\documentclass[prd,aps,reprint,10pt,nofootinbib]{revtex4-1}

\usepackage{amsmath, hyperref}
\usepackage{amssymb}
\usepackage{mathrsfs}
\usepackage{graphicx}
\usepackage{color}

\begin{document}
	
\title{Quantum approach to a Bianchi I singularity}

 \author{Ana Alonso-Serrano} 
 \email{ana.alonso.serrano@aei.mpg.de}
 \affiliation{Max-Planck-Institut f\"ur Gravitationsphysik 	(Albert-Einstein-Institut)\\ Am M\"uhlenberg 1, 14476 Potsdam, Germany}
 
 \author{Martin Bojowald}
 \email{bojowald@gravity.psu.edu}
 \affiliation{Institute for Gravitation and the Cosmos,
 	The Pennsylvania State
 	University,\\
 	104 Davey Lab, University Park, PA 16802, USA}

\author{David Brizuela}
\email{david.brizuela@ehu.eus}
\affiliation{Fisika Teorikoa eta Zientziaren Historia Saila, UPV/EHU, 	644 P.K., 48080 Bilbao, Spain}

\date{\today}

\begin{abstract}
The approach of a quantum state to a cosmological
singularity is studied through the evolution of its moments in a simple
version of a Bianchi I model. In spite of the simplicity, the model exhibits
several instructive and unexpected features of the moments. They are
investigated here both analytically and numerically in an approximation in
which anisotropy is assumed to vary slowly, while numerical methods are also
used to analyze the case of a rapidly evolving anisotropy. Physical conclusions are drawn
mainly regarding two questions. First, quantum uncertainty of anisotropies
does not necessarily eliminate the existence of isotropic solutions, with
potential implications for the interpretation of minisuperspace truncations as
well as structure-formation scenarios in the early universe. Secondly,
back-reaction of moments on basic expectation values is found to delay the
approach to the classical singularity.
\end{abstract}


\maketitle

\section{Introduction}

The dynamics of anisotropic cosmological models are believed to give a
reliable description of the approach to a space-like singularity in general
relativity, based on the Belinskii--Khalatnikov--Lifshitz (BKL) \cite{BKL}
scenario. It is therefore of interest to analyze in detail the behavior of
quantized anisotropic models in order to determine whether a singularity may
persist in quantum gravity. The most generic dynamics, given by the Bianchi
IX model, can be rather complicated classically \cite{Billiards}, but
even in this case it consists of long stretches of time during which the
dynamics resembles that of the simpler Bianchi I model.

The main goal of this paper is to analyze how the presence of anisotropies may
affect the behavior of a quantum state, parametrized by its quantum
fluctuations and higher-order moments. These parameters can be considered
coordinates of a quantum phase space that extends the classical phase space of
the volume and anisotropy degrees of freedom, parametrized here in a
Misner-like fashion. The quantum parameters, as opposed to a wave function,
preserve the geometrical nature of the classical phase-space problem and are
therefore appropriate for a quantum understanding of the BKL scenario.

Misner variables \cite{Misner,Mixmaster} describe a homogeneous geometry not
directly through the coefficients in a line element but rather through the
volume and two anisotropy (or shape) parameters. We will further restrict the
dynamics by assuming that only one of the anisotropy parameters is
non-zero. As geometrical variables, we will therefore have the volume, one
anisotropy parameter, the moments of each of these variables and their
momenta, and cross-moments between volume and anisotropy. Even in the
restricted setting of a single anisotropy parameter and a quantum dynamics
truncated to some fixed moment order, the parameter space is therefore rather
large, making the analysis non-trivial and instructive.

For generic anisotropy, the system of dynamical equations for moments is
highly coupled and hard to solve analytically. We will therefore introduce an
approximation in which anisotropy varies much more slowly than the volume, in
which case several analytical expressions can be obtained. Numerical results
for moments up to fifth order are shown for generic anisotropy.

In addition to computational questions in the analysis of our system, this
paper highlights two kinds of physical interpretations of the technical
results. First, isotropic models within anisotropic ones can be used as test
systems of the minisuperspace truncation, in which the relation between a
symmetric quantum model and a less-symmetric one is an important open
question; see for instance \cite{MiniValid}. Our equations will allow us to
determine conditions on the moments of a state in the anisotropic model such
that it follows the behavior of the isotropic model. A general argument
against minisuperspace truncations is that quantum uncertainty relations
prevent anisotropic degrees of freedom from being completely absent,
questioning the validity of a quantum model in which those degrees of freedom
have been neglected. We will find that in our model, on the contrary, it is
possible to find states that follow exactly isotropic behavior. 

While this result may be considered supportive of minisuperspace truncations
at least in the types of models studied here, it also strengthens questions
that have been raised about quantum scenarios of structure formation
\cite{InflStruc,CQCFieldsHom}: In early-universe cosmology, inhomogeneity is
supposed to be generated out of quantum fluctuations of an initially
homogeneous state, but if the dynamics is translation invariant, it should
preserve the homogeneity of any initial state. In our case, similarly, the
isotropy of an initial state is preserved by quantum evolution, but only under
additional conditions on higher-order moments.

Our second application is about the behavior of a quantum state approaching an
anisotropic singularity. We find that different kinds of moments play
different roles. We therefore determine which moments can be used as
indicators of singular behavior. Such results are useful for establishing the
genericness of various proposals to avoid singularities by quantum
effects. Often, such proposals are analyzed by using a specific class of
initial or evolving states. While we also fix our initial states, making the
common Gaussian choice, we are able to track the moments that grow most
strongly and might therefore have a dominant effect on the quantum behavior
near a singularity. We also draw lessons about possible modifications of the
approach to a singularity, which seems to be slowed down by back-reaction at
least with respect to the time variable chosen here, given by
deparametrization with respect to a scalar field.

\section{Canonical description of the classical model}

The metric of an anisotropic Bianchi I universe is given by,
\[
 {\rm d}s^2=-N^2 {\rm
  d}t^2+\sum_{k=1}^{3}a_k^2{\rm d}x_k^2\,,
\]
where $a_k$ are the scale factors in the different spatial directions, and $N$
is the lapse function.  In the variables introduced by Misner
\cite{Misner,Mixmaster}, this line element takes the form
\begin{eqnarray}
{\rm d}s^2=-N^2 {\rm d}t^2+e^{2\alpha} \sum_{k=1}^3 e^{2\beta_k}
{\rm d}x_k^2\,.
\end{eqnarray}
The spatial volume is described by the variable $\alpha$, defined by
$e^\alpha:=(a_1a_2a_3)^{1/3}$, whereas the three shape-parameters
$\beta_k=\ln (a_k/e^{\alpha})$ measure the degree of anisotropy of each
spatial direction. These three variables are not independent but
satisfy the constraint $\beta_1+\beta_2+\beta_3=0$.  Therefore, for
convenience, we construct two independent shape-parameters defined as
\begin{eqnarray}
\beta_+&:=&-\frac{1}{2}\beta_3=-\frac{1}{2}\ln (a_3/(a_1a_2a_3)^{1/3}), \nonumber \\
\beta_-&:=&\frac{1}{2\sqrt{3}}(\beta_1-\beta_2)=\frac{1}{2\sqrt{3}}\ln (a_1/a_2)\,.
\end{eqnarray}

We will consider a free, massless scalar field $\phi$ as the matter source,
with conjugate momentum $p_{\phi}$. The Hamiltonian constraint is then given by
\begin{eqnarray} \label{C1}
{\cal C}:= e^{-3\alpha}(-p_\alpha^2+p_+^2+p_-^2)+\frac{1}{2}e^{-3\alpha}p_\phi^2=0,
\end{eqnarray}
where we have absorbed constant factors (such as Newton's constant) in
$p_{\phi}$.  Using this constraint, the conjugate momenta of the configuration
variables $(\alpha, \beta_{\pm},\phi)$ are obtained in terms of their
derivatives with respect to coordinate time $t$,
\begin{equation}
p_\alpha=-\frac{e^{3\alpha}}{2 N}\alpha_{,t},\qquad
p_{\pm}=\frac{e^{3\alpha}}{2N}\beta_{\pm,t},\qquad
p_\phi=\frac{e^{3\alpha}}{N}\phi_{,t}\,.  
\end{equation}

In order to simplify the effects of the anisotropy on the system, we will
consider only one anisotropic direction by choosing a vanishing $\beta_-$.  In
this way, the directions $x_1$ and $x_2$ will be isotropic, as their
corresponding scale factors are equal, $a_1=a_2$, whereas the direction $x_3$
will generically be anisotropic. Therefore, we have just one shape-parameter,
$\beta:=\beta_+$, which measures the ratio of the scale factor $a_3$ with
respect to the geometric mean of the three scale factors. Alternatively, we
could eliminate the matter content and use $\beta_-$ as internal time, such
that $\frac{1}{2}p_{\phi}^2$ in the following expressions would play the role
of $p_-^2$. Our results therefore apply to a matter model with restricted
anisotropy, or to a vacuum model with full anisotropy. While different
deparametrization choices lead to equivalent classical results, they do
not always imply equivalent quantum corrections. In what follows we will
consider only one specific deparametrization in order to obtain a specific system of equations that
determines the dynamics of quantum states.

Instead of choosing the logarithm of the scale factor $\alpha$ as our basic
variable, we will use the spatial volume $v=e^{3\alpha}$.\footnote{More precisely, we will
assume that $v$, as a phase-space variable, can take both signs in order to
obtain a simple phase space. The definition $v=e^{3\alpha}>0$ then describes
one set of solutions but not the entire phase space. This distinction will
briefly be relevant below, when we introduce a suitable quantum
representation.}
Its conjugate
momentum is proportional to the Hubble parameter, describing the isotropic
rate of expansion of the universe:
\begin{equation}\label{defpv}
p_v=\frac{1}{3}e^{-3\alpha}p_\alpha=-\frac{2 v_{,t}}{ N v}\,.
\end{equation}
This variable is preferred for numerical purposes because it places the
singularity at a finite value of the geometric variable, $v=0$. Moreover, even
though the constraint 
\begin{eqnarray}\label{constraint}
{\cal C}=\frac{ 1}{v}\left(-v^2p_v^2+p_\beta^2+\frac{1}{2}p_\phi^2\right)=0
\end{eqnarray}
in the volume parameter may appear more complicated than the original
(\ref{C1}), it will be straightforward to interpret the quantum dynamics of
moments of the volume, as opposed to moments of its logarithm.

Since neither $\phi$ nor $\beta$ appear explicitly in the constraint, their
conjugate momenta $p_\phi$ and $p_\beta$ are conserved quantitities.  The
expression of the Hamiltonian constraint $\cal C$ is also conserved through
evolution, thus one can infer that the combination $v p_v$ is another constant
of motion, which will appear throughout this paper.  As can be seen in the
definition \eqref{defpv}, this combination is proportional to the momentum
$p_\alpha$, but we will refer to it as $\gamma:=vp_v$ because we will not
consider it a basic canonical variable.

In fact,
after performing the deparametrization with respect to $\phi$, $\gamma$ will
represent the unconstrained Hamiltonian $H_{\rm
  iso}=-p_{\phi}|_{p_{\beta}=0}$ of the reference isotropic model \eqref{isotropich}.  Including
anisotropy, the deparametrized dynamics is generated by the Hamiltonian
\begin{equation}\label{classicalH}
H:=-p_{\phi}=(\gamma^2-p_\beta^2)^{1/2}\,,
\end{equation}
implying the classical equations of motion
\begin{eqnarray}
\dot{v}&=&\frac{\gamma}{H}v,\\
\dot{p}_v&=&-\frac{\gamma}{H}p_v,\\
\dot{\beta}&=&-\frac{p_\beta}{H},\\
\dot{p}_\beta&=&0,
\end{eqnarray}
where the dot represents a derivative with respect to the scalar field $\phi$.
Since the equations are symmetric under the transformation $p_v\rightarrow
-p_v$, $p_\beta\rightarrow -p_\beta$, and $\phi\rightarrow-\phi$ and we are
interested in an expanding universe with a singularity in the past, towards
decreasing $\phi$, we will without loss of generality choose a positive sign
for both $p_\beta$ and $p_v$ (and, thus, also for $\gamma$).  In this way, and
with the choice of sign taken for $p_\phi$ when solving the constraint
\eqref{constraint}, the universe expands as $\phi$ increases and the
singularity is located at $\phi\rightarrow-\infty$.

The canonical variables of the system are $(v,p_v,\beta,p_\beta)$. An
analysis of the structure of the classical model shows that a canonical
transformation to $\gamma$ and its conjugate $\ln v$ would simplify the
canonical quantization of the system. However, such a non-linear
transformation would imply a complicated mapping between moments that does not
preserve the semiclassical order. The physical interpretation of quantum
moments would then be obscured because the meaning of a quantum fluctuation of
$\ln v$ is not as clear as the volume fluctuation of $v$ itself.

The equations of motion can easily be solved:
\begin{eqnarray}\label{classicalsol1}
v(\phi)&=&v(0) \exp{\left(\frac{\gamma}{H} \phi\right)},
\\\label{classicalsol2}
p_v(\phi)&=&p_v(0) \exp{\left(-\frac{\gamma}{H} \phi\right)}, \\\label{classicalsol3}
\beta(\phi)&=&\beta(0)-\frac{p_\beta}{H}\phi,
\end{eqnarray}
while $p_{\beta}$ is a constant of motion, as already seen.  Here,
$v(0),p_v(0)$ and $\beta(0)$ are initial values of the different variables at
$\phi=0$. As the volume $v$ tends to zero, approaching the singularity located
at $\phi\rightarrow-\infty$, its conjugate momentum, $p_v$, diverges
exponentially, keeping their product $\gamma$ constant. The ratio $\gamma/H$
parametrizes the rate of collapse of the volume towards the singularity. On
the other hand, the shape-parameter $\beta$ increases as a linear function of
$\phi$ with a velocity controlled by the constant of motion $p_\beta$, making
the universe more and more anisotropic as it approaches the singularity.  The
variable $\beta$ tends to (plus) infinity for $\phi\to-\infty$, producing a
singularity as the scale factor $a_3$ tends to zero.  The other two scale
factors, $a_1$ and $a_2$, which are equal in our restricted model, may be
non-zero, but such that $v=a_1a_2a_3\to0$ for $\phi\to-\infty$. Using the
defining relationships of our variables and the solutions
\eqref{classicalsol1}--\eqref{classicalsol3}, we can write
\begin{eqnarray}
  a_1(\phi)&=& \!v(\phi)^{1/3} \exp(\beta(\phi)/2) \nonumber \\\nonumber &=& v(0)^{1/3} \exp(\beta(0)/2)
  \exp((\gamma/3-p_{\beta}/2)\phi/H).
\end{eqnarray}
Therefore, $a_1$ approaches zero or $+\infty$ at the singularity, depending on
the sign of $(\gamma/3-p_{\beta}/2)$. In the vacuum model, we would have
$p_{\phi}=0$ and therefore $\gamma=p_{\beta}$, such that $a_1\propto
\exp(-p_{\beta}\phi/(6H))\to\infty$ at the usual Kasner singularity. With
scalar matter, however, $p_{\beta}$ is a free parameter restricted only by
$p_{\beta}\leq \gamma$. This condition does not fix the sign of
$\gamma/3-p_{\beta}/2$, and $a_1$ may approach zero or $+\infty$ depending on
the initial conditions.

Let us remark that this condition introduces certain boundaries in the phase
space of the system. Nonetheless, if the initial conditions are given inside
these boundaries, the system will never cross them as the Hamiltonian is
conserved throughout evolution. Note that $|p_{\beta}|>|\gamma|$ is not
possible on the constraint surface defined by (\ref{constraint}). Therefore, any initial
state that fulfills this inequality would not be physical. For the quantization
procedure, in order to obtain a well-defined Hamiltonian, one can simply replace
the classical expression (\ref{classicalH}) with $|\gamma^2-p_{\beta}^2|^{1/2}$.
Nevertheless, we will not spell
this out explicitly because it is not relevant for moment equations.

\section{Quantum dynamics}

Having the classical dynamics of the system under control, we proceed to
analyze its quantum dynamics following a formalism based on a moment
decomposition of the wave function developed for quantum cosmology in
\cite{EffAc,HigherMoments}. The quantum dynamics of this model is ruled by a
Hamiltonian $\hat H$, that depends on the basic operators $\hat v$, $\hat
p_v$, $\hat\beta$, and $\hat p_\beta$.\footnote{Note that
we define our
classical phase space such that $v$ is the oriented volume and therefore
can take both signs. The phase space is therefore a standard cotangent
bundle of the plane and can be quantized by standard means, with self-adjoint
basic operators $\hat{v}$, $\hat{p}_v$, $\hat{\beta}$ and $\hat{p}_{\beta}$.}
In order to analyze the quantum
evolution produced by this Hamiltonian, we will define the following moments,
which encode the complete information of the quantum state,
\begin{equation}\label{defmoments}
G^{ijkl}:= \langle (\hat v-v)^i (\hat p_v-p_v)^j (\hat \beta-\beta)^k (\hat
p_\beta-p_\beta)^l \rangle_{\rm Weyl}\,,
\end{equation}
where the subscript ``Weyl'' indicates totally symmetric ordering of the
operators, and the expectation values $v:=\langle \hat v \rangle$,
$p_v:=\langle \hat p_v \rangle$, $\beta:=\langle \hat\beta \rangle$ and
$p_\beta:=\langle \hat p_\beta \rangle$ have been defined. We will refer to
the sum of the indices of a given moment $i+j+k+l$ as its {\em order}. This
definition will be relevant later on when we consider truncations of the
system. 

Unlike the basic expectation values, moments of a state are not completely
arbitrary but restricted by (generalized) uncertainty relations which follow
from the positivity condition of an algebraic state
(the derivation of such generalized inequalities
for the case of one degree of freedom is studied in
\cite{Bri15}). These restrictions will
play an important role in some of our discussions, but in specific cases we
will mainly refer to the well-known second-order version, which is nothing but
Heisenberg's uncertainty relation.
Provided these general conditions are obeyed by a
given set of moments, a state with these moments does exist. However, it is
not guaranteed to be a pure state, demonstrating the general nature of states
included in the parametrization by moments.

\subsection{Effective Hamiltonian and equations of motion}

In this subsection we will present the effective Hamiltonian
that rules the dynamics of the quantum moments. Their equations of motion will
be derived and the structure of the corresponding system of equations will be
discussed.  Following this analysis we will perform a redefinition of our
variables, in particular the relative moments (\ref{newvariables}) will be defined, in
order to simplify the coupling between different equations.

The dynamics of these variables is given by the following effective
Hamiltonian, defined as the expectation value of the quantum
Hamiltonian operator, which is assumed to be Weyl-ordered:
\begin{eqnarray}\nonumber
H_Q:&=&\langle \hat H(\hat v,\hat p_v,\hat p_\beta) \rangle
\\\label{qhamiltonian}
&=&
\sum_{i,j,k}\frac{1}{i!j!k!}\frac{\partial^{i+j+k}H(v,p_v,p_\beta)}{\partial 
  v^i\partial p_v^j\partial p_\beta^k} G^{ij0k},
\end{eqnarray}
where $H(v,p_v,\beta)$ is the classical Hamiltonian \eqref{classicalH} and the
sum runs over all non-negative integer values of $i,j$ and $k$. (If a
Hamiltonian operator with a different ordering is preferred, the effective
Hamiltonian would contain terms explicitly depending on $\hbar$ that result from
re-ordering operations.)  In particular, if $i=j=k=0$, then $G^{ij0k}=1$
because the state is normalized. The corresponding term in the sum therefore
produces the classical Hamiltonian, $H(v,p_v,p_{\beta})$, evaluated in the
basic expectation values.  For instance, the second-order Hamiltonian is
\begin{widetext}
\begin{eqnarray} 
	H_Q\!=\!H\!-\!\frac{v^2 p_v^2}{2H^3} G^{0 0 0 2}
	+\frac{ v p_v}{H^3}\left(H^2-p_\beta^2\right)G^{1 1 0 0}
	+\frac{p_\beta}{H^3} v p_v^2 G^{1 0 0 1}
	-\frac{v^2p_\beta^2}{2 H^3} G^{0 2 0 0}
	-\frac{p_v^2p_\beta^2}{2 H^3} G^{2 0 0 0}
	+\frac{p_\beta }{H^3} v^2 p_v G^{0 1 0 1}\!. \label{HSecond}
\end{eqnarray}
\end{widetext}

Since $\beta$ does not appear in
the classical Hamiltonian, only moments unrelated to $\beta$ (and thus of the
form $G^{ij0k}$) appear in the expression of the effective Hamiltonian. This
fact implies that $p_\beta$, as well as all its pure fluctuations (moments of
the form $G^{000i}$), are constants of motion for the quantum
dynamics. Nonetheless, $\gamma=vp_v$ will not be a constant of motion
at the quantum level because the full quantum Hamiltonian $H_Q$
\eqref{qhamiltonian}, but not the classical Hamiltonian $H$
\eqref{classicalH}, is conserved by quantum evolution. Here, we
define $\gamma$ as the product of expectation values of $\hat{v}$ and
$\hat{p}_v$. An alternative definition, using 
\begin{equation}
\tilde{\gamma}= \frac{1}{2}\langle\hat{v}\hat{p}_v+\hat{p}_v\hat{v}\rangle=
vp_v+G^{1100}\,, 
\end{equation}
is less convenient for our purposes. If one were to use $\hat{\gamma}$ as a
basic operator, as done in affine quantum cosmology
\cite{AffineQG,AffineSmooth,AffineSing,SpectralAffine,MixAffine},
$\langle\hat{\gamma}\rangle$ would be conserved as a consequence of
$[\hat{\gamma},\hat{H}]=0$. However, neither $\gamma$ nor $\tilde{\gamma}$
need be conserved in our system because the assumed Weyl ordering in $\hat{v}$
and $\hat{p}_v$ as basic operators implies that an operator quantizing a
classical expression, that depends on $v$ and $p_v$ only through $\gamma$, is
not required to depend on $\hat{v}$ and $\hat{p}_v$ only through
$\frac{1}{2}\langle\hat{v}\hat{p}_v+\hat{p}_v\hat{v}\rangle$. We will see
explicit solutions in which, indeed, neither $\gamma$ nor $\tilde{\gamma}$ are
conserved.

The equations of motion for the different variables are obtained by computing
Poisson brackets with the effective Hamiltonian. The Poisson brackets between
two expectation values are related to the expectation value of their
commutator by the relation 
\begin{equation}\label{Bracket}
 \{\langle \hat X\rangle,\langle \hat Y\rangle\} =-i \langle[\hat X,\hat
 Y]\rangle\,, 
\end{equation}
extended to products of expectation values by the Leibniz rule.  This
expression is standard for basic expectation values, while it defines an
extension of the classical bracket for moments. A general expression for the
brackets of moments is known in closed form \cite{EffAc,HigherMoments}, but it
is rather lengthy and will not be displayed here. (See also
\cite{Bosonize,EffPotRealize} for the structure of the underlying Poisson
manifold.)  These brackets are not canonical and they contain linear and
quadratic terms in moments.\footnote{Even for canonical pairs of basic operators, such as $\hat{x}$ 
and $\hat{p}$ in quantum mechanics, the brackets of moments are non-canonical. For instance, 
$\{G^{02},G^{20}\}=4G^{11}$ is not constant, and therefore not canonical.}
In particular, the origin of the linear terms lies
in the reordering of operators and, therefore, they appear multiplied by
certain power of $\hbar$. Each of this $\hbar$ factors is considered as
increasing the total moment-order by two.  Our main arguments will use the
schematic form of the bracket,
\begin{equation} \label{GG}
 \{G^{ijkl},G^{mnop}\}\sim GG^{qrst},
\end{equation}
where ``$GG$'' on the right-hand side represents a finite sum of terms
quadratic in moments (or a moment multiplied by certain power of $\hbar$) of a
total order $(qrst)$ such that $q+r+s+t=i+j+k+l+m+n+o+p-2$.  This general
statement about orders follows from an application of (\ref{Bracket}), in
which the commutator always reduces the total moment order by two.

In this way one can, for instance, obtain the equation of motion for the
volume:
\begin{eqnarray}\nonumber
\dot v &=&\{v,H_Q\}
\\
&=&\frac{\gamma}{H}
v+\sum_{i,j,k}\frac{1}{i!j!k!}\frac{\partial^{i+j+k+1}H(v,p_v,p_\beta)}{\partial
  v^i\partial p_v^{j+1}\partial p_\beta^k} G^{ij0k}\,, 
\end{eqnarray}
where we have used the fact \cite{EffAc} that all the expectation values (in
particular $v$) Poisson commute with the moments \eqref{defmoments}.  One can
then proceed in this way to find the equations of motion for all the
variables.  

In general, the equations of motion for the moments and expectation values
form a highly coupled infinite system of equations. Therefore, one usually
needs to implement a truncation in order to solve them. The main assumption is
that for semiclassical states peaked around a classical trajectory,
there is a hierarchy of moments ruled by their order.  For such states,
higher-order moments are then less relevant than lower-order moments. In
particular, for the numerical solutions that will be performed later on, we
will consider the system of equations up to fifth order in moments.  The
equations of motion up to such a high order are much too lengthy to be
displayed here. Hence, in order to give a grasp of the system we are
dealing with, all the equations up to second order in moments are displayed in
Appendix~\ref{app_eq2G}.  In addition, in Appendix~\ref{app_eq5} the
evolution equation for the volume is given, truncated at fifth order.

For the specific Hamiltonian \eqref{classicalH} under consideration, the
equations are not completely coupled.  In particular, since only moments
$G^{ij0k}$ unrelated to the shape-parameter appear in the Hamiltonian, the set
of equations of motion for the variables $\{v, p_v, \beta, p_\beta,
G^{ij0k}\}$ forms an independent subsystem of equations that can be solved on
its own. This is due to the fact that the Poisson bracket $\{G^{ij0l},
G^{kl0m}\}$, which must be computed to obtain the evolution equation for
$G^{ij0l}$, does not generate any moment of the form $G^{abcd}$, with $c\neq
0$. In fact, the equation of motion for $\beta$ depends on $\{v, p_v, p_\beta,
G^{ij0k}\}$ but not the other way around, and one can thus conclude that the
system $\{v, p_v, p_\beta, G^{ij0k}\}$ is independent of the rest.  One can
even remove the dependence on the volume from this system by performing a
further change of variables, as shown below.

For the main analysis of this paper, instead of using the absolute moments
$G^{ijkl}$, we will use the relative moments
\begin{equation} \label{newvariables}
K^{ijkl}:=\frac{G^{ijkl}}{v^i p_v^j}=\gamma^{-j} v^{j-i} G^{ijkl}.
\end{equation}
Furthermore, the momentum of the volume, $p_v$, will be replaced by the
isotropic Hamiltonian $\gamma:=v p_v$.  As will be explained in
Sec.~\ref{sec_quasiharmonic}, a convenient property of this new set of
variables $\{v,\gamma,\beta,p_\beta,K^{ijkl}\}$ is that all but the volume $v$
are constants of motion in the limit of a slowly-evolving anisotropy
$(p_\beta\ll \gamma)$.  More importantly, with this new set of variables the
couplings between different equations of motion simplify considerably as the
equations of motion for the variables $\{\gamma, p_\beta,K^{ij0l}\}$ decouple
from the equations for $v$, $\beta$ and the rest of the moments, $K^{ijkl}$
with $k\neq 0$.  In fact, the volume $v$ and the shape-parameter $\beta$ only
appear explicitly in their own equations of motion as a time derivative (or a
logarithmic derivative in the case of the volume). Schematically one
can write the equations of motion as 
\begin{eqnarray} \label{schematic-eqs1}
\frac{\dot{v}}{v}&=&F_1(\gamma, p_\beta, K^{ij0l}),\\
\dot{\beta}&=&F_2(\gamma, p_\beta, K^{ij0l}),\\
\dot{\gamma} &=&F_3(\gamma, p_\beta, K^{ij0l}),\\
\dot{p}_\beta&=&0,\\
\dot{K}^{ab0d}&=&F_{ab0d}(\gamma, p_\beta, K^{ij0l})
\end{eqnarray}
where the right-hand sides are given in terms of the constant $p_\beta$,
the Hamiltonian of the reference isotropic model $\gamma$,
and moments $K^{ij0l}$ unrelated to the shape-parameter,
but are independent of the volume $v$ and $\beta$.
Therefore, in order to obtain the quantum back-reaction effects
on the classical trajectories,
it is enough to consider this subsystem of equations.
Similarly, the equation of motion for a generic
moment has the form
\begin{equation} \label{schematic-eqs2}
 \dot{K}^{abcd}=F_{abcd}(\gamma, p_\beta, K^{ijkl})\,.
\end{equation}
Hence, the dynamics of the moments is only affected by the expectation values
$\gamma$ and $p_\beta$, but not by $v$ and $\beta$. The explicit form
of this system of equations, truncated at second order in moments, is
shown in Appendix \ref{app_eq2K}.

Finally, as with the classical system, the equations are symmetric under the
transformation $p_v\rightarrow -p_v$, $\gamma\rightarrow -\gamma$,
$p_\beta\rightarrow -p_\beta$, and $\phi\rightarrow-\phi$, provided that the
moments are also transformed as $G^{ijkl}\rightarrow (-1)^{j+l} G^{ijkl}$ or,
equivalently, $K^{ijkl}\rightarrow (-1)^l K^{ijkl}$.  Therefore, as already
commented above, a positive sign for $\gamma$ and $p_\beta$ will be considered
throughout the paper.

\subsection{The isotropic (harmonic) case} \label{sec_harmonic}

Using the basic variables defined here, the isotropic case is formally
recovered by choosing $\beta$ and $p_\beta$ to vanish, along with all the
moments with some contribution from the anisotropic sector (that is,
$G^{ijkl}$ with $k+l\neq 0$). Restricting all moments of this form is not
consistent with uncertainty relations in the anisotropy sector. The
restriction therefore amounts to a minisuperspace truncation of isotropic
geometries within anisotropic (but still homogeneous) ones. In principle,
therefore, the reduction is not expected to define a subset of quantum
solutions in the anisotropic model. A detailed analysis of solutions will
nevertheless show that isotropic solutions do exist within the anisotropic
quantum model.

The classical Hamiltonian (\ref{classicalH}) is then simplified to be a linear
function of $v$ and $p_v$,
\begin{equation}\label{isotropich}
H^{\rm iso}=|\gamma|=|v p_v|\,.
\end{equation} 
{}From the perspective of the quantum dynamics,
this case is very special, as the Hamiltonian turns out to be
\emph{harmonic}. (It is quadratic in phase-space variables. A linear canonical
transformation maps it to an inverted harmonic oscillator.)
The most important property of this kind of Hamiltonians is that different
orders in moments are  not coupled to one another. Furthermore, the classical
equations of motion do 
not get 
corrections 
by quantum moments; there is no quantum back-reaction. Therefore, the
expectation values $v$ and $p_v$ 
follow exactly their classical trajectories \eqref{classicalsol1}--\eqref{classicalsol2},
\begin{equation}\label{harmonicevolution}
v=v(0)e^\phi,\qquad p_v=p_v(0)e^{-\phi}\,.
\end{equation}

In addition, it is easy to obtain and solve the equations of motion for the
quantum moments.  Note that the infinite sum that defines the effective
Hamiltonian \eqref{qhamiltonian} is reduced to a finite sum, as only second-order derivatives are
nonvanishing. Therefore, for this harmonic case one obtains the following
quantum Hamiltonian,
\begin{equation}
H_Q^{\rm iso}=v p_v+G^{1100}
\end{equation}
without any truncation. The equations of motion for the different moments,
\begin{equation}\label{eqharmonicG}
\dot{G}^{ij00}=(i-j)G^{ij00}\,,
\end{equation}
indeed shows that there is no coupling between the equations of motion for
different orders. Solving this equation, one obtains an exponential evolution
for the moments,
\begin{equation}
G^{ij00}(\phi)=G^{ij00}(0) e^{(i-j)\phi}\,.
\end{equation}

In summary, as one approaches the singularity at $v=0$, moments $G^{ij00}$
with $i>j$ decrease exponentially, whereas moments $G^{ij00}$ with $i<j$
follow an exponentially increasing behavior.  Finally, moments of the form
$G^{ii00}$ are constants of motion.  Taking into account the time dependence
for the expectation values \eqref{harmonicevolution}, we note also that all
relative moments $K^{ij00}$ \eqref{newvariables} are constant throughout
evolution since the time dependence of the absolute moments $G^{ij00}$ is
compensated for by that of the expectation values.

\section{\mbox{Slowly-evolving anisotropy ($p_\beta\ll \gamma$)}}\label{sec_quasiharmonic}

A natural generalization of the isotropic case analyzed in the
previous subsection is given by the case with a slowly-evolving
anisotropy. In this section, we will analyze such a case, and will present the
analytical form of the evolution of the moments and expectation values. In
addition to providing a detailed analytical understanding, this case will
serve as a reference to analyze more generic cases numerically.

As noted after equation \eqref{classicalsol3}, $p_\beta$ is the velocity of the
shape-parameter $\beta$ and therefore measures the rate of (an)isotropization
of the universe, while $\gamma$ is a measure of the velocity of expansion of
the isotropic reference model. Therefore, the system dynamics should be close
to the isotropic dynamics whenever $p_\beta\ll \gamma$ is obeyed so that
$H\approx H^{\rm iso}$. This condition means that the evolution of the
anisotropy is slow compared with the rate of expansion of the volume. These
are statements about the rates of change rather than the size of the
homogeneous region. The approximation may therefore be used in the late
universe (where the homogeneous volume may be assumed macroscopic) or in the
early universe close to a spacelike singularity (where the BKL scenario
suggests the existence of microscopic homogeneous patches).

In this section we will consider an expansion of the system of equations for
large values of the parameter $\gamma$.  In particular, the equations of
motion for the different moments take the form
\begin{equation}
\dot{G}^{ijkl}=(i-j)G^{ijkl}+{\cal O}(\gamma^{-1})\,.
\end{equation}
We therefore recover similar equations as in the harmonic case above
\eqref{eqharmonicG}, but for all the moments and not only for moments
$G^{ij00}$ of the isotropic sector.  It is easy to see that the generator of
these equations is the effective Hamiltonian $H=v p_v+{\cal O}(\gamma^{-1})$.
These equations can be solved right away,
\begin{equation} \label{isomoment}
G^{ijkl}(\phi)=G^{ijkl}(0) e^{(i-j)\phi}\,.
\end{equation}

As one would expect, in the case of slowly-evolving anisotropy the dynamics is
dominated by the isotropic sector. In particular, the increasing or
decreasing behavior of a corresponding moment $G^{ijkl}$ is completely
determined by the difference between its isotropic indices $i$ and
$j$. Furthermore, at this level of approximation, the equations for the
expectation values do not get any quantum back-reaction effects from the
moments, and thus expectation values follow the classical
trajectory. In this way, the relative moments $K^{ijkl}$ \eqref{newvariables}
are conserved quantitities,
\begin{equation}
K^{ijkl}=K^{ijkl}(0)\,.
\end{equation}

Let us now analyze the behavior of the system at next order in $1/\gamma$. We
will consider an expansion for large $\gamma$, keeping the volume and the
relative moments $K^{ijkl}$ constant. The classical Hamiltonian takes the
form
\begin{equation}
H= \gamma\bigg( 1 -\frac{p_\beta^2}{2 \gamma^2}\bigg)+{\cal
  O}(p_\beta^4/\gamma^3)\,. 
\end{equation}
It can be expanded in order to get the quantum Hamiltonian,
\begin{widetext}
\begin{equation}\label{approxHq}
H_Q= \gamma(1+K^{1100})-
\frac{1}{2\gamma}\sum_{i,j} (-1)^{i+j}\bigg( p_\beta^2 K^{ij00}+
2 p_\beta K^{ij01}+K^{ij02}\bigg)
+{\cal O}(3),
\end{equation}
\end{widetext}
where the sum runs over all non-negative integer values of $i$ and $j$.
In this expansion, ${\cal O}(3)$ stands for terms of the form
$K^{ij0k}p_\beta^{n-k}/\gamma^{n-1}$ for $n\geq 3$ and $0\geq k\geq n$.
Therefore, the present approximation should be valid
as long as all those terms are small. Here, we first expanded the classical
Hamiltonian and then derived its effective expression. It is easy to see that
the order can be reversed without changing the result, for instance using the
second-order example (\ref{HSecond}).

In the classical Hamiltonian one can define the dimensionless anharmonicity
parameter $p_\beta/\gamma$, which is a constant of motion and measures the
departure of the system from the harmonic behavior. In the quantum system,
however, there are infinitely many more parameters that produce an anharmonic
behavior, such as the moments $K^{ij02}$ that appear explicitly in the
Hamiltonian above \eqref{approxHq} and might generate an anisotropy even if
$p_\beta=0$ and $\beta=0$ at some initial time.  In fact, this case will be
analyzed in detail in the next section.

The equations of motion for the expectation values,
generated by the approximate Hamiltonian \eqref{approxHq},
are
\begin{eqnarray}\label{approxeqv}
\dot{v}&=& v
,\\
\dot{\gamma}&=&-\frac{1}{2 \gamma}\sum_{i,j} (-1)^{i+j}(i-j)\nonumber \\ && \times \bigg( p_\beta^2 K^{ij00}+
2 p_\beta K^{ij01}+K^{ij02}\bigg), \\ \label{approxeqbeta}
\dot{\beta}&=&-
\frac{1}{\gamma}\sum_{i,j} (-1)^{i+j}\bigg( p_\beta K^{ij00}+K^{ij01}\bigg)\,.
\end{eqnarray}
At this level of approximation, all moments of the form $K^{ij0k}$ are
constants of motion since the Poisson brackets $\{K^{ij0k},K^{lm0n}\}/\gamma$
are of the form $KK^{ab0d}/\gamma^2$ with $a=i+j-1$, $b=l+m-1$ and $d=k+n$,
while ``$KK$'' is interpreted as explained for ``$GG$'' in (\ref{GG}).  Any
such term turns out to be of order ${\cal O}(3)$ and, thus, should be
neglected. In particular, this includes all \emph{purely isotropic moments}
of the form $K^{ij00}$, which only involve the isotropic
variables. Furthermore, all the moments that appear in the Hamiltonian
\eqref{approxHq} and in the equations for the expectation values above
\eqref{approxeqv}--\eqref{approxeqbeta}, are also included in this
category. Therefore, these last equations can be easily integrated to obtain
the evolution of the expectation values,
\begin{eqnarray}\label{approxvsol}
v&=&v_0 e^{\phi},\\\label{approxgammasol}
\gamma&=&\sqrt{\gamma_0^2+r \phi},\\\label{approxbetasol}
\beta&=&\beta_0-\frac{2\widetilde p_\beta}{r} \sqrt{\gamma_0^2+r\phi},
\end{eqnarray}
where the following constants have been defined:
\begin{eqnarray}\label{rconstant}
r& :=&-\sum_{i,j} (-1)^{i+j}(i-j)\nonumber \\
&& \times \bigg( p_\beta^2 K^{ij00}+
2 p_\beta K^{ij01}+K^{ij02}\bigg),\\
\widetilde p_\beta& :=&\sum_{i,j} (-1)^{i+j}\bigg( p_\beta
K^{ij00}+K^{ij01}\bigg) \nonumber \\
&=&p_\beta+\sum_{i\geq 1,j\geq 1}\bigg( p_\beta
K^{ij00}+K^{ij01}\bigg)\,. \label{pbtilde}
\end{eqnarray}
(For these generic solutions, we have assumed that $r\not=0$; see below.)
Note that, in general, neither $\gamma=vp_v$ nor
$\tilde{\gamma}=vp_v+G^{1100}$ are constant, in contrast to the classical
solution. The dynamics of $\gamma$ is instead governed by the constant $r$,
which is purely quantum and vanishes in the classical limit. Since we are
assuming a large value of $\gamma_0$, the solutions
\eqref{approxgammasol}--\eqref{approxbetasol} can be approximated by linear
functions:
\begin{eqnarray}
\gamma\approx \gamma_0+\frac{r}{2\gamma_0}\phi,\\\label{approxbeta}
\beta\approx \beta_0-\frac{\widetilde p_\beta}{\gamma_0}\phi.
\end{eqnarray}

The solution \eqref{approxbetasol} for the shape-parameter is valid
only for the generic case $r\neq 0$. For special states in the quantum case,
$r=0$ is compatible with uncertainty relations. For instance, at second order,
while $r$ depends on the moments $K^{2000}$ and $K^{0200}$, which cannot both
be zero, it does so in an antisymmetric way because of the factor of $(i-j)$ in
(\ref{rconstant}). Provided $K^{2000}=K^{0200}$, these moments therefore
cancel out. Moreover, while the general expression for $r$ depends on
$K^{ij02}$, the second-order moment $K^{0002}$ does not contribute because of
the same factor of $(i-j)$.  In the special case of $r=0$, then,
$\gamma=\gamma_0$ is constant while $\beta$ changes linearly with time as
in the approximate solution \eqref{approxbeta},
\begin{equation}
  \beta= \beta_0-\frac{\widetilde p_\beta}{\gamma_0}\phi\,.
\end{equation}
Therefore, as in the classical limit, $\beta$ is a linear function of $\phi$,
but with a regularized value of $p_\beta$ which takes into account quantum
effects.

The moments, except for the constant $K^{ij0k}$ which still follow their
isotropic behavior, do feel the effects of anisotropy and are no longer
constant.  One class of moments --- those that imply only
one factor in the shape-parameter $K^{ij1k}$ --- have simple equations of
motion since they only contain constant
moments of the form $K^{nm0l}$. Schematically, the equations for such moments
are given as
\begin{equation}
\dot{K}^{ij1k}=\frac{c_{ijk}}{\gamma},
\end{equation}
with certain constants $c_{ijk}$ that depend on $p_\beta$ and
moments of the form $K^{nm0l}$. This equation can be integrated, which
gives rise to
\begin{equation}
K^{ij1k}=d_{ijk}+\frac{2 c_{ijk} \sqrt{\gamma_0^2+r \phi }}{r},
\end{equation}
with integration constants $d_{ijk}$ for $r\neq 0$, and
\begin{equation}
K^{ij1k}=d_{ijk}+\frac{c_{ijk}}{\gamma_0}\phi,
\end{equation}
for $r=0$.

This pattern continues, allowing us to iteratively solve for the behavior of
all moments. In the next step, using (\ref{GG}), $\dot{K}^{ij2k}$ is given by
a sum of terms of the form $K^{ij1(k-1)}/\gamma$, each of which has a time
dependence $d_{ij(k-1)}(\gamma_0^2+r\phi)^{-1/2}+2c_{ij(k-1)}/r$ for
$r\not=0$.  Integrating, we have
\begin{equation}
 K^{ij2k}= \frac{c_{ijk}^{(2)}\phi}{r}+
 d_{ijk}^{(2)}\frac{\sqrt{\gamma_0^2+r\phi}}{r}+e_{ijk}^{(2)}
\end{equation}
with new constants $c_{ijk}^{(2)}$, $d_{ijk}^{(2)}$ and $e_{ijk}^{(2)}$.  The
dominant behavior is linear in $\phi$.  Finally, it is possible to obtain
that, at this level of approximation, a general moment $K^{ijnk}$ has the
dominant behavior
\begin{equation} \label{Kn}
 K^{ijnk}\sim \frac{(\gamma_0^2+r\phi)^{n/2}}{r^n}.
\end{equation}
The demonstration follows by induction. Note that $\dot{K}^{ij(n+1)k}$ is a
sum of terms of the form $KK^{ijn(k-1)}/\gamma$ (again, see (\ref{GG}) for the
meaning of ``$KK$''), which all have the dominant behavior
\begin{equation}
 \dot{K}^{ij(n+1)k} \sim \frac{(\gamma_0^2+r\phi)^{(n-1)/2}}{r^n},
\end{equation}
according to (\ref{Kn}). Integrating this expression, it is then
straightforward to obtain the form (\ref{Kn}) for $n+1$.  For
third-order moments, this result is confirmed in Appendix~\ref{app_sol}.  For
the particular case $r=0$, in which (\ref{Kn}) no longer applies, the
evolution of the moments is faster and a moment of the form $K^{ijkl}$ is
given by a polynomial of order $k$ in $\phi$.

In summary, for this quasiharmonic case, we have found that up to order ${\cal
  O}(2)$ the volume follows its classical trajectory, whereas $\gamma$ is not
constant anymore but is a linear function in $\phi$. The shape-parameter
$\beta$ is a linear function of $\phi$, but with a quantum-corrected
slope. Finally, depending on whether the constant $r$ is vanishing or not,
relative moments $K^{ijkl}$ go either as $\phi^k$ or as
$\phi^{k/2}$. Therefore, their index on $\beta$ governs their
evolution rate.

\section{On the quantum generation of anisotropy}

Before analyzing numerically generic values of $p_\beta$, let us look at the
particular case of $\beta=0$ and $p_\beta=0$. Classicaly there is then no
initial anisotropy and the spacetime will remain isotropic throughout
evolution. In a quantized model, however, one would expect that some
anisotropy is generated by quantum fluctuations (or certain higher moments)
which are constrained by uncertainty relations to be non-zero. This
expectation is a common criticism of minisuperspace quantizations, which start
with symmetry reductions at the classical level and therefore ignore
fluctuations of non-symmetric variables. While symmetry reduction leads to
special solutions of the classical theory, it is not clear whether their
minisuperspace quantizations can be considered approximations of solutions of
some full theory of quantum gravity.

In a more specific context, it would be interesting if non-symmetric degrees
of freedom could, in fact, be generated by quantum effects. This possibility,
as a physical scenario, is usually considered for inhomogeneity rather than
anisotropy in order to explain structure formation in the early universe. In
this context, it would be desirable to excite non-symmetric degrees of freedom
even if the initial state is symmetric (such as the homogeneous vacuum). Our
model can be used as a test system in which inhomogeneity is replaced by more
tractable anisotropy.

We will therefore be interested in initial states with vanishing
$p_{\beta}$. Nevertheless, the discussion in the present section goes beyond
what we found in the preceding section because we will assume that some
anharmonicity parameters of the form $K^{ij0k}p_\beta^{n-k}/\gamma^{n-1}$ are
not negligible for certain $n$, such that $0\geq k\geq n$.

In order to make the appearance of anisotropy transparent, we begin by
analyzing the equation of motion for the shape-parameter, $\beta$. 
If $p_\beta=0$, it takes the particularly simple form
\begin{equation}\label{dotbeta}
\dot{\beta}=\sum_{i,j,k} \frac{1}{\gamma^k} f(i,j,k) K^{ij0(2k+1)},
\end{equation}
where the sum runs over all non-negative integer values of $i$, $j$ and $k$,
and $f$ is a function that only depends on the indices $i$, $j$ and
$k$. Therefore, the moments that produce a nonvanishing derivative for the
shape-parameter are precisely those that are unrelated to this variable and,
moreover, have odd order in $p_\beta$.  For any such moment, uncertainty
relations do not imply any lower bound. Therefore, it is consistent to assume
that all $K^{ij0(2k+1)}$ are zero in a certain class of states. Specific
examples can easily be constructed using products of Gaussian wave functions,
such that $K^{ij0(2k+1)}=K^{ij00}K^{000(2k+1)}=0$ because all odd-order
moments vanish for a Gaussian.

It is therefore possible to choose an initial state for which the right-hand
side of equation \eqref{dotbeta} is zero. As time goes on, the dynamics might
activate some of the relevant moments, $K^{ij0(2k+1)}$, in which case the time
derivative of the shape-parameter would become non-zero and an anisotropy
would be generated. However, using the detailed dynamics at least up to fifth
order in moments, we have analytically confirmed that the time derivative of any
$\dot{K}^{ij0(2k+1)}=0$ is identically zero provided $p_{\beta}=0$ and all
$K^{ij0(2k+1)}=0$ at an initial time.  The right-hand side of \eqref{dotbeta}
is then vanishing at all times, and no anisotropy is generated. 

This result can be understood based on the general behavior of moment
equations. Since the effective Hamiltonian does not depend on $\beta$ or its
moments, a non-zero $\{K^{ij0(2k+1)},H_Q\}$ can be obtained only via the
$(v,p_v)$-part of the moments. Moreover, $H$ is a function of $p_{\beta}^2$,
such that only even-order moments of the form $K^{mn0(2l)}$ contribute to
$H_Q$ when $p_{\beta}=0$. The first property implies that $p_{\beta}$-orders
of moments add up in $\{K^{ij0(2k+1)},H_Q\}$, which then contains only moments
of odd order in $p_{\beta}$ based on the second property. Therefore, if all
$K^{ij0(2k+1)}$ are zero initially, they remain zero if $p_{\beta}=0$.

Although this result follows directly from properties of the moment brackets,
it is somewhat unexpected based on general arguments about limitations of
minisuperspace quantization. Our result, however, relies on the specific
dynamics of the moments and not just on general expectations on implications
of uncertainty relations. It is also consistent with detailed studies made in
the case of inhomogeneity \cite{InflStruc, CQCFieldsHom}, where an expectation
opposite to the usual criticism of minisuperspace quantization has been
formulated.  Nevertheless, since our result relies on the detailed dynamics,
it may well change if other models are considered, for instance those with a
non-vanishing anisotropy potential. In our model, the coupling between the
different degrees of freedom is not strong. It could therefore be possible
that our result comes about because the Hamiltonian \eqref{classicalH} is a
function (a square root) of the harmonic (free) Hamiltonian $(v^2
p_v^2-p_\beta^2)$, where the different sectors $(v,p_v)$ and $(\beta,p_\beta)$
are completely decoupled.\footnote{Required inequalities such as 
$v^2 p^2_v - p^2_\beta \geq 0$ can be imposed on initial values and therefore do 
not introduce dynamical coupling terms.}

At high orders in moments, the nonlinearities and strong couplings make it
difficult to obtain analytical solutions. We have, however, been able to
obtain the general solution of the system up to third order in moments. At
this order, the evolution of the expectation value of the volume $v$ and $\gamma$ are given by, 
\begin{eqnarray} \label{vgamma}
v&=&v_0 \left| \gamma^2+r \phi\right|^{-(K^{0  1  0  2}+\frac{1}{2}K^{1  0  0 
		2}-\frac{1}{2}K^{0  0  0  2})/r}e^{\phi}, \nonumber \\ 
\gamma&=&\sqrt{\gamma_0^2+r \phi } ,
\end{eqnarray}
where $r=K^{1002}-K^{0102}$ (here assumed to be non-zero) is the truncation of
\eqref{rconstant} to third order. The shape-parameter has the following form,
\begin{equation}
\beta=\beta_0+\frac{K^{0  0  0  3}}{r\gamma}-\frac{2 \widetilde p_\beta}{r}\gamma ,
\end{equation}
in terms of $\gamma$ from (\ref{vgamma}),
where 
\begin{equation} \label{pbtilde3}
\widetilde
p_\beta=(K^{1 0 0 1}+K^{0 1 0 1}+K^{1 1 0 1}+K^{2 0 0 1}+K^{0 2 0 1}) ,
\end{equation}
is the truncation of (\ref{pbtilde}) to third order.

All moments that appear in our solutions for the expectation values are
constants of motion.  In general a moment of the form $K^{ijkl}$ is a
polynomial of order $k$ in $\gamma$ and therefore changes like $\phi^{k/2}$
for large $\phi$.  Moreover, the second-order moments that involve either $v$
or $p_v$ depend on $\phi$ through an expression logarithmic in $\gamma$. There
is no such logarithmic term in third-order moments, except for $K^{0120}$ and
$K^{1020}$, in which case this term is multiplied by $\hbar^2$.  Therefore, in
addition to the momentum $p_\beta$ and its pure fluctuations, $K^{000i}$,
almost all the moments $K^{ij0l}$ unrelated to the shape-parameter $\beta$ are
constants of motion. The only three nonconstant fluctuations of the isotropic
sector increase as logarithmic functions of $\gamma$,
\begin{eqnarray}
K^{0 2 0 0}&=&b_{1}+ \left(1+\frac{K^{0  1  0  2}+K^{1  0  0  2}}{K^{0  1  0 
		2}-K^{1  0  0  2}}\right) \ln\gamma ,\\ 
K^{1 1 0 0}&=& b_{2}-\ln\gamma ,\\
K^{2 0 0 0}&=&b_{3}+\left(1-\frac{K^{0  1  0  2}+K^{1  0  0  2}}{K^{0  1  0 
		2}-K^{1  0  0  2}}\right) \ln\gamma\   ,
\end{eqnarray}
where $b_1,b_2,b_3$ are real constants.  The explicit form for the rest of the
moments is given in Appendix \ref{app_solpbeta0}, including fluctuations of
the anisotropic sector and different correlations between the two sectors.

The generic solution presented above is not valid in the particular case in
which the moments $K^{0102}$ and $K^{1002}$ are equal, such that
$r=0$. If $r=0$, $\gamma$ is constant and the
volume depends on $\phi$ by
\begin{equation}
v=v_0 \exp \left[\phi+\left(2 K^{0  0  0  2}-3 (K^{0  1  0  2}+K^{1  0  0 
	2})\right)\frac{\phi}{4 \gamma ^2}\right] ,  
\end{equation}
while the shape-parameter increases as a linear
function in $\phi$, as in the classical case,
\begin{equation}
\beta = \beta_0-\frac{2 \gamma ^2
	\widetilde p_\beta+K^{0  0  0  3}}{2 \gamma ^3}\phi\ .
\end{equation}
Therefore, even if $\beta_0$ is vanishing, the quantum moments will produce an
anisotropy by acting as an effective $p_\beta$, unless $K^{0003}$ and those
that appear in (\ref{pbtilde3}) vanish. Nonetheless, as commented above,
this is allowed by uncertainty relations and one can indeed choose initial
states that will never generate an anisotropy.

The moments that were constant in the previous generic solution are
also constant in the case of $r=0$, as well as the isotropic correlation
$K^{1100}=\tilde b_{2}$. Therefore, in the isotropic sector,
only pure fluctuations of $v$ and $p_v$ are dynamical, and they increase
faster (as linear functions of $\phi$) than in the previous case:
\begin{eqnarray}
K^{0200}&=& \tilde b_{1}-\frac{K^{0  1  0  2}+K^{1  0  0  2}}{2 \gamma ^2} \phi, \nonumber \\
K^{2 0 0 0}&=& \frac{K^{0  1  0  2}+K^{1  0  0  2}}{2 \gamma ^2} \phi +\tilde b_{3}.
\end{eqnarray}
The rest of the moments are explicitly given in Appendix~\ref{app_solpbeta0}.

Our model therefore suggests some middle ground between the pessimistic
expectations formulated in the two distinct contexts of minisuperspace
quantization on one hand, and structure formation on the other. General
criticism of the minisuperspace quantization argues that none of the solutions
of a minisuperspace model are relevant for the full dynamics because
non-symmetric degrees of freedom will always get excited, while concerns about
structure formation are based on the statement that a symmetric initial state
cannot evolve into a structured state in which the symmetry is broken. In our
model, we find that anisotropy is, generically, generated, but there are also
states that retain an initial isotropic form. The latter is not prohibited by
uncertainty relations.

\section{Numerical analysis of the model}

The complicated structure of equations of motion at high orders in moments is
illustrated by the equations collected in the appendices.  It implies that
analytical investigations are possible only in certain particular cases, as
shown here for exact isotropy or slowly varying anisotropy.  Going beyond
these regimes requires a numerical implementation to solve the equations of
motion and interpret the dynamics. For this numerical study, a truncation of
the system to fifth order in moments has been considered by neglecting sixth
and higher-order moments. (For the consistency of such truncations, see
\cite{Counting}.) 

The full phase space of our system has coordinates $\{v,
\gamma,\beta,p_\beta,K^{ijkl}\}$. As shown by the schematic equations
\eqref{schematic-eqs1}--\eqref{schematic-eqs2}, the equations of motion for
$\{\gamma,p_\beta,K^{ij0k}\}$ form an independent subsystem that is decoupled
from the remaining equations.  Therefore, one can solve this subsystem without
considering the whole set of equations of motion.  Once the evolution of
$\{\gamma,p_\beta,K^{ij0k}\}$ has been obtained, the equations of motion for
the rest of the moments, $K^{ijkl}$ with $k\neq 0$, and for the expectation
values $v$ and $\beta$ can be solved.

For specific numerical solutions, we will assume an initial quantum state
given by a product of two Gaussians, one in the volume $v$ and the other one
in the shape-parameter $\beta$, centered at initial expectation values $v_0$
and $\beta_0$ respectively.  The moments for such a state are

\begin{widetext}
\begin{eqnarray}
K^{ijkl}=
\left\{
\begin{array}{cl}
\gamma_0^{-j} v_0^{j-i}\,  2^{-(i+j+k+l)}
\hbar^{j+l}\sigma_v^{i-j}\sigma_\beta^{k-l}\frac{i!\,j!}{\left(i/2\right)!\left(j/2\right)!} 
\frac{k!\,l!}{\left(k/2\right)!\left(l/2\right)!},& \mbox{if all indices are
  even,} 
\\&
\\
0,&
\mbox{otherwise,}
\end{array}
\right.
\end{eqnarray}
\end{widetext}
where $\sigma_v$ and $\sigma_\beta$ are the Gaussian widths in the volume and
in the shape-parameter, respectively, and $\gamma_0$ is the initial value of
$\gamma$.

In order to construct a semiclassical state peaked on a classical trajectory,
we will impose small initial relative fluctuations. In
particular, for the isotropic sector, both $K^{2000}$ and
$K^{0200}$ have to be small:
\begin{equation}
K^{2000}=\frac{\sigma_v^2}{2 v^2_0}\ll 1,\qquad  K^{0200}=\frac{v_0^2\hbar^2}{2
\sigma_v^2 \gamma_0^2}\ll 1\,. 
\end{equation}
Therefore, the Gaussian width needs to be chosen as $\hbar v_0/\gamma_0\ll
\sigma_v\ll v_0$. In addition, if one requires the state to be unsqueezed,
with equal absolute fluctuations for both conjugate variables,
$G^{0200}=G^{2000}$, then one gets the specific value $\sigma_v=\sqrt{\hbar}$
for the width. This is the value we will consider for both the Gaussian width
in the volume $\sigma_v$ and in the shape-parameter
$\sigma_\beta=\sqrt{\hbar}$. From this point on, and for all the numerical
simulations, $\hbar$ will be set equal to one. In these units, the Gaussian
widths will then be chosen as $\sigma_v=\sigma_\beta=1$, and the requirement
of an initial peaked state is summarized by the condition $1\ll
v_0\ll\gamma_0$. However, the values of the
Gaussian widths have also been altered by several orders of magnitudes to
check that this choice does not qualitatively affect our main results.

Regarding initial conditions for the expectation values
$\{v_0,\gamma_0,\beta_0, p_{\beta} \}$, following the discussion of peaked
states, $v_0$ and $\gamma_0$ have been chosen very large, obeying the
constraint $v_0\ll \gamma_0$.  In the anisotropic sector, $\beta$ does not
appear in the equations of motion for the moments and is therefore less
relevant than the other variables, while we would like to analyze the behavior
of the system for different values of $p_\beta$. For convenience, we will choose a small
initial value for $\beta_0$ (around unity), so that we begin our simulations
with a nearly isotropic universe. Due to the form of the
Hamiltonian \eqref{classicalH}, the maximum allowed value for $p_\beta$ is
$p_\beta=\gamma$. Therefore, we will explore the behavior of the system for
different values of $p_{\beta}$ between a very small value (corresponding to
slowly varying anisotropy) and the fixed $\gamma_0$. Since we allow for small
values of both $\beta_0$ and $p_{\beta}$, we do not use a
sharply peaked state in these variables because relative moments in the
anisotropy sector may be large if the basic expectation values are small.

Based on the possible values of $p_{\beta}$, this section is divided into
two parts. We will first consider slowly varying anisotropy, that is
$p_\beta\ll \gamma$, in order to test the analytical results obtained in
Sections~\ref{sec_harmonic} and \ref{sec_quasiharmonic}.  In the second part
we will consider solutions in which the shape-parameter is evolving more
rapidly, leaving the previous quasi-harmonic regime. In this case, $p_\beta$
will be of the same order of magnitude as $\gamma$. Since we observe different
behaviors of the system for different values of $\gamma_0$, we will further
subdivide the second part into three parts.  In the first and second parts we
will analyze the evolution of the moments for small and big values of
back-reaction on the evolution of the expectation values $v$ and $\beta$
will be studied.

\subsection{Slowly varying anisotropy ($p_\beta\ll \gamma$)}

In Section \ref{sec_quasiharmonic}, we have found approximate analytical
solutions for slowly-evolving anisotropy.  At zeroth order in the
anharmonicity parameters, all moments $K^{ijkl}$ are constants of motion,
whereas at next order, including terms of order $1/\gamma$, their evolution is
determined either by a polynomial of order $k$ in internal time $\phi$ ($r=0$)
or by a dependence of the form $\phi^{k/2}$ ($r\not=0$).

Numerically we have observed that moments of the form $K^{ij00}$ follow the
same qualitative behavior as the isotropic ones: They are constant throughout
the whole evolution and do not feel the presence of an anisotropy as long as
$p_\beta$ is small.  Moments of the anisotropy sector, $K^{00ij}$, do not have
an isotropic counterpart. According to our analytical results, pure moments of
$p_\beta$, $K^{000n}$, are exactly conserved during evolution, which is easily
confirmed numerically. Perhaps surprisingly, we find out that pure
fluctuations of $\beta$, $K^{00n0}$, are also conserved up to a high degree of
precision.  Based on the approximate analytical solution, by contrast, one
would expect an evolution of the form $\phi^n$ or $\phi^{n/2}$. This
discrepancy seems to be a consequence of the specific initial state, in
particular the uncorrelated nature of the anisotropic state for $\beta$ and
$p_{\beta}$. Therefore, most correlations of the form $K^{ijnm}$, with
$n\neq 0$ and $m\neq 0$, are zero; and they are the only moments that contribute to
$\dot{K}^{00n0}$. The correlations are not conserved and may therefore build
up during evolution, but the rate is suppressed by a factor of $1/H$ compared
with the evolution of expectation values.

While generic
correlations of the form $K^{00nm}$, with $n\neq 0$ and $m\neq 0$, are not
conserved in numerical solutions but rather evolve as linear functions in time
$\phi$, they do not follow the analytical behavior (\ref{Kn}) unless $n=2$.

\begin{figure}[]
	\begin{center}
		\includegraphics[scale=0.30]{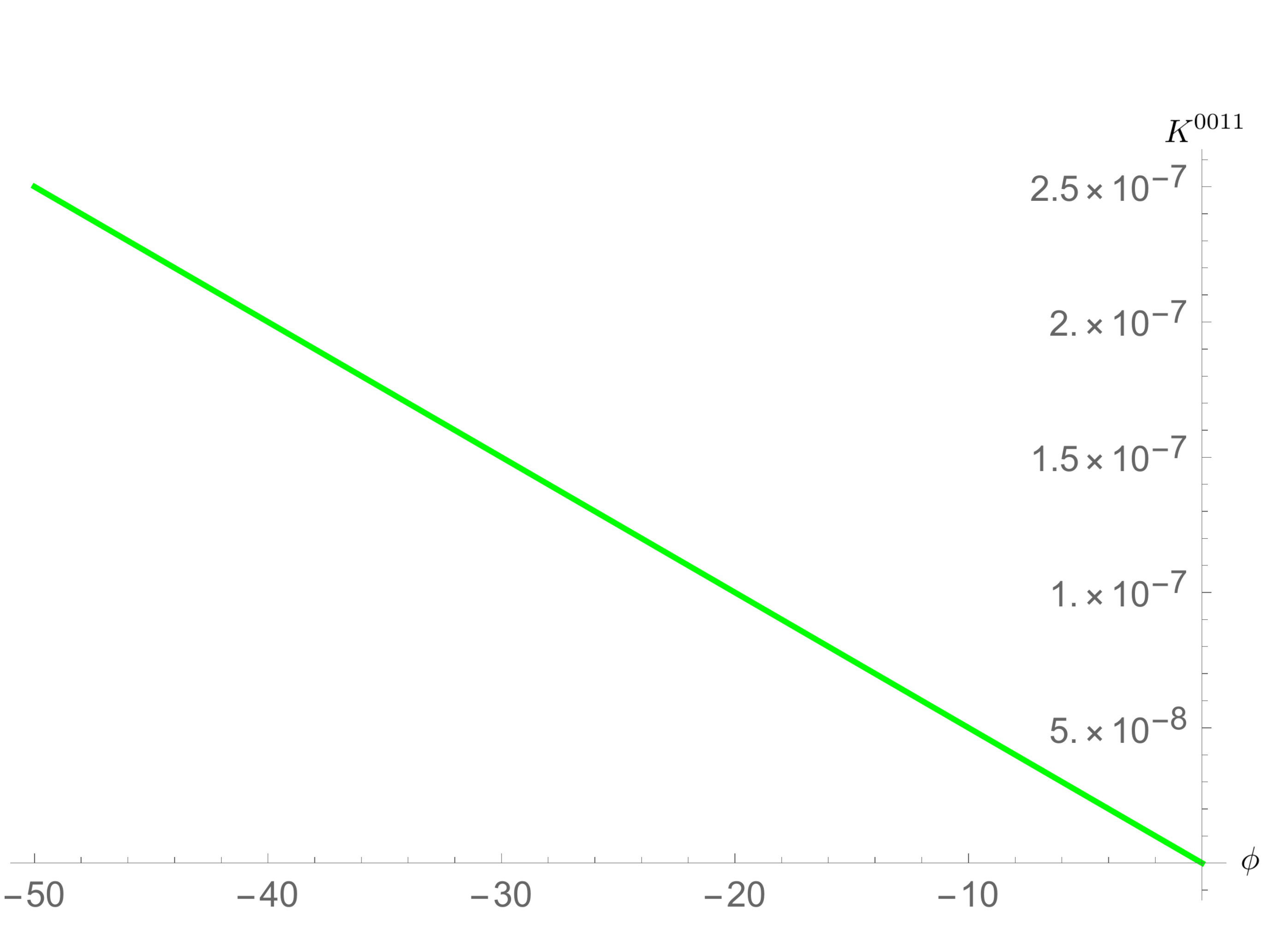}
		\caption{This plot shows an example of the evolution of a purely anisotropic moment, more precisely
the correlation between the shape-parameter and its conjugate momentum $K^{0011}$,
for the case of a slowly-variying anisotropy. For this plot $\gamma_0=10^8$ and $p_\beta=10^{-1}$ have been chosen.}
		\label{fig1}
	\end{center}
\end{figure}

\begin{figure}[]
	\begin{center}
		\includegraphics[scale=0.30]{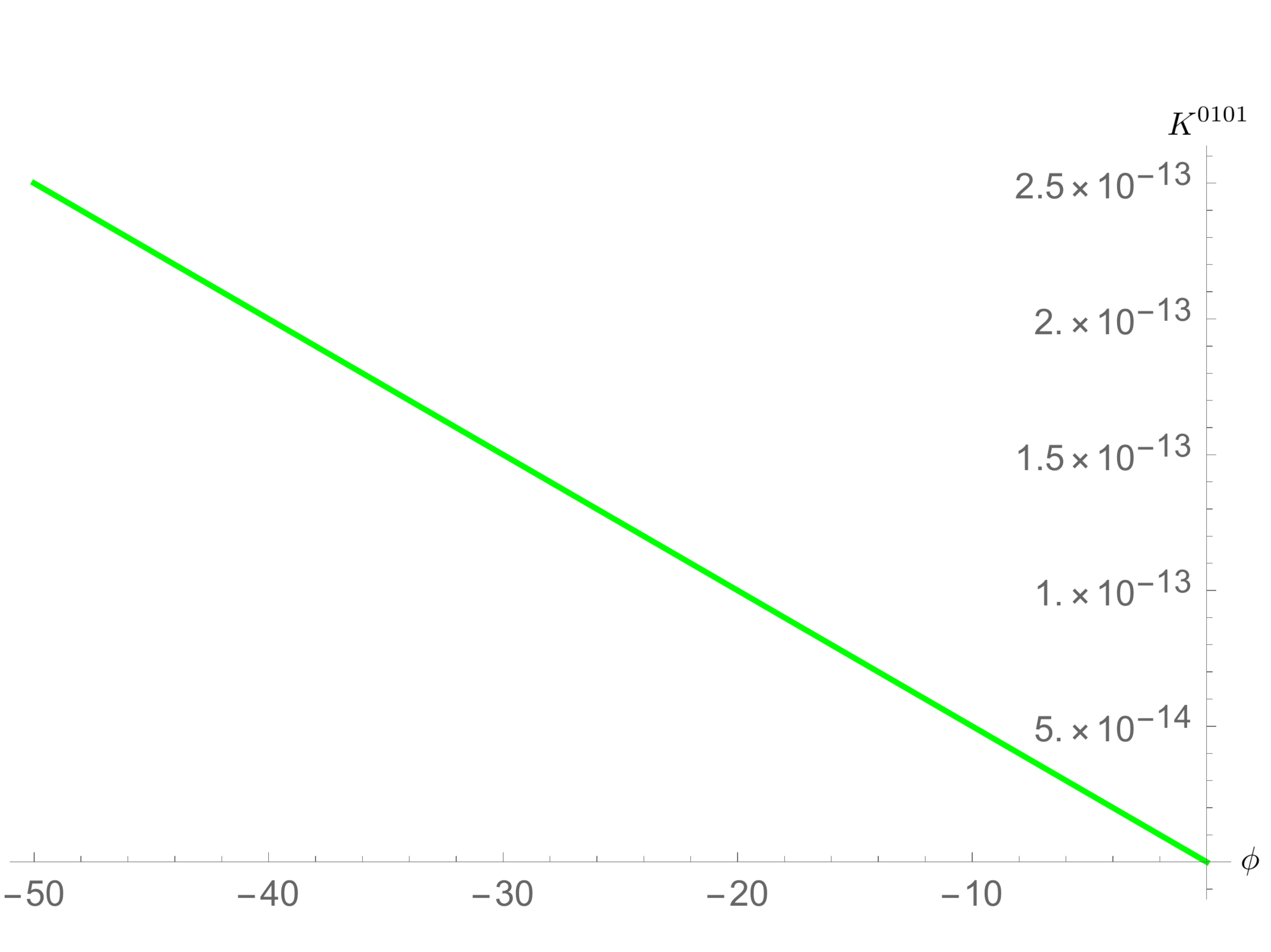}
		\includegraphics[scale=0.30]{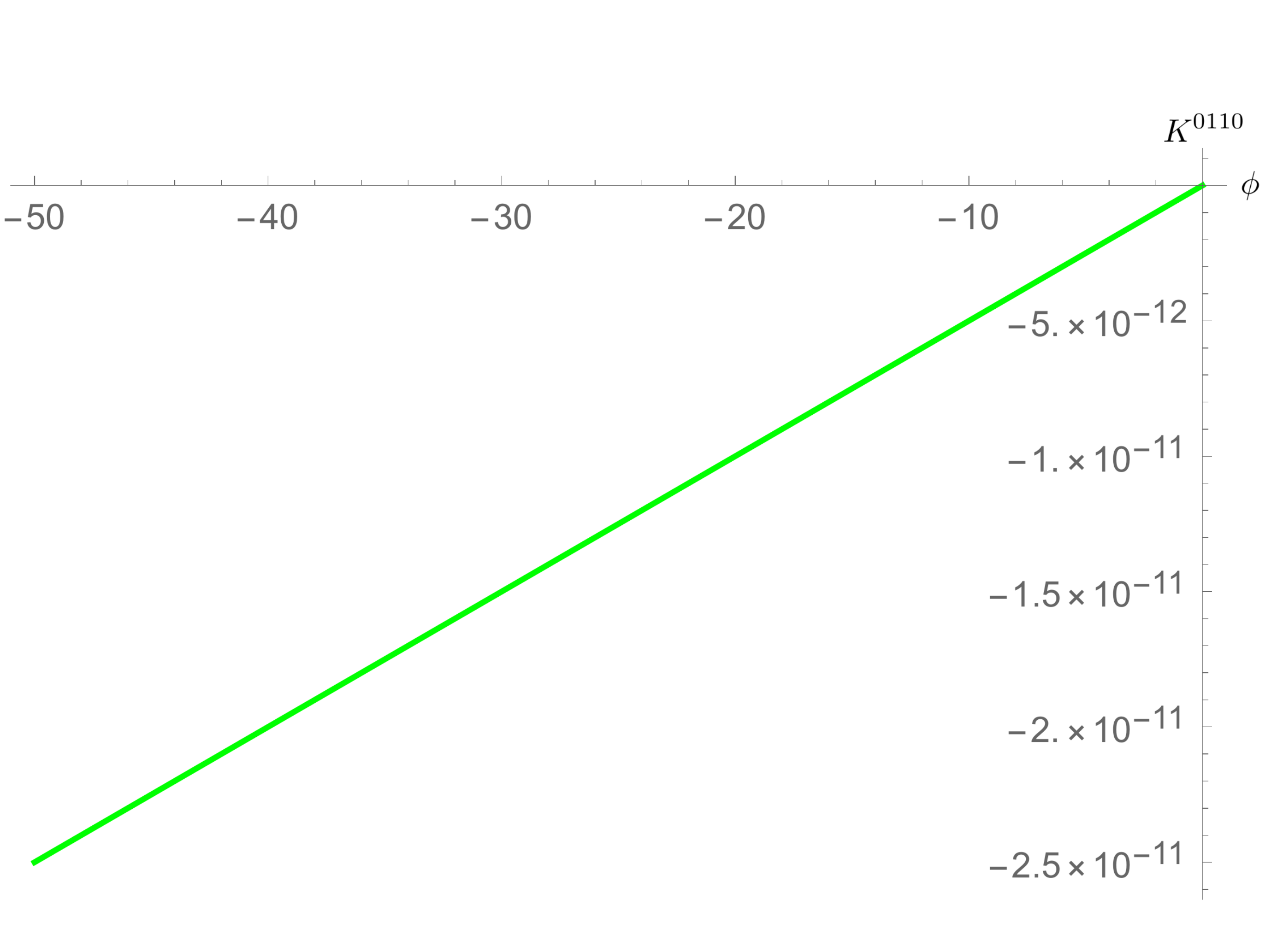}
		\caption{Here we show the linear evolution of two moments, $K^{0101}$ and $K^{0110}$, for the
slowly varying anisotropy case. Even if both have a common isotropic counterpart, as their
$v$ and $p_v$ indices are the same, they exhibit two opposite behaviors: $K^{0101}$ is increasing,
whereas $K^{0110}$ is decreasing. But, in module, both go to infinity and thus they slightly accelerate the
exponential diverging of the corresponding absolute moments, $G^{0101}$ and $G^{0110}$.
In any case, this variation is very small as compared with the dominant isotropic behavior.
As in the previous figure, here we have chosen $\gamma_0=10^8$ and $p_\beta=10^{-1}$.
		}
		\label{fig2}
	\end{center}
\end{figure}

\begin{figure}[]
	\begin{center}
		\includegraphics[scale=0.30]{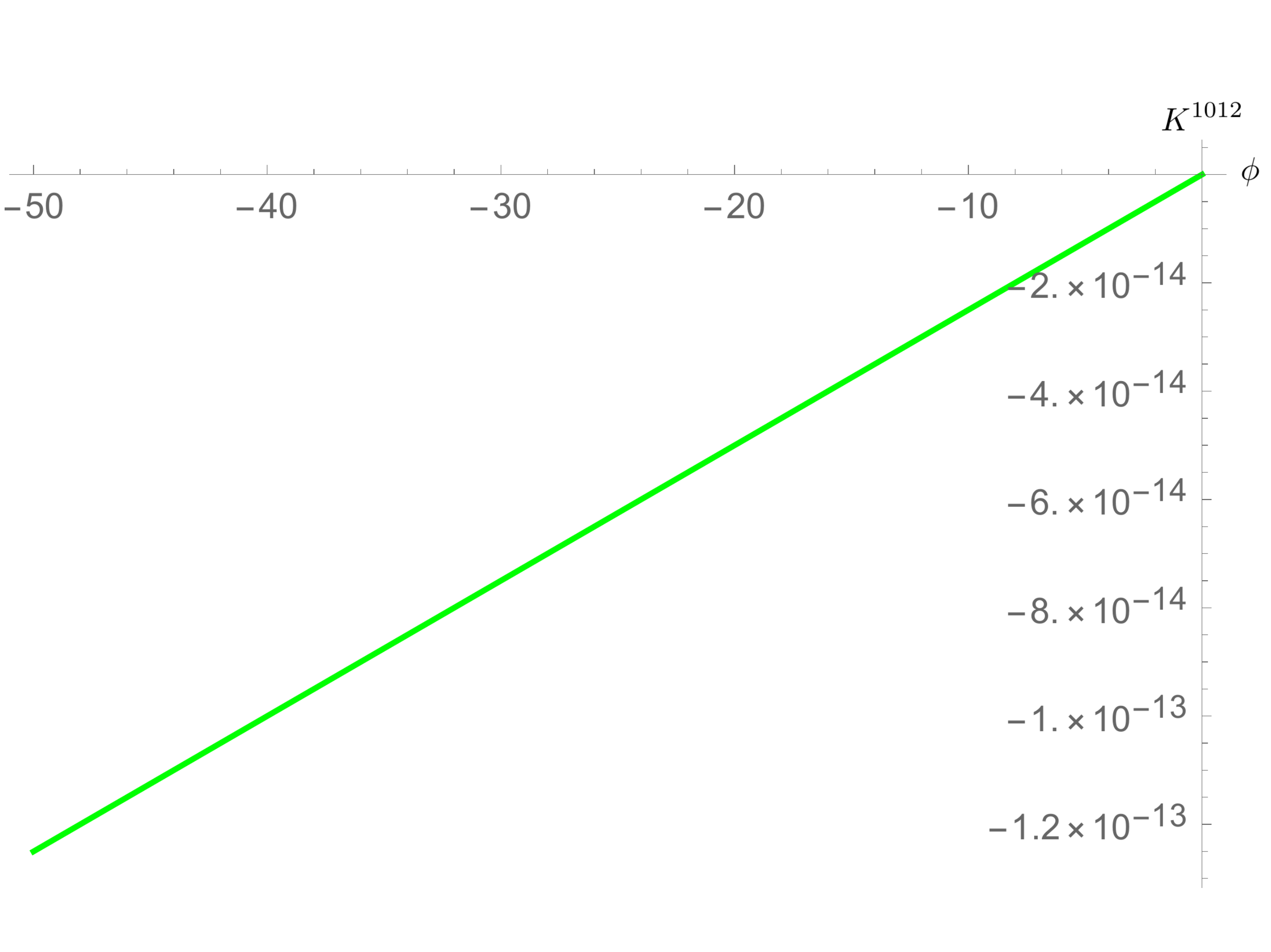}
		\includegraphics[scale=0.30]{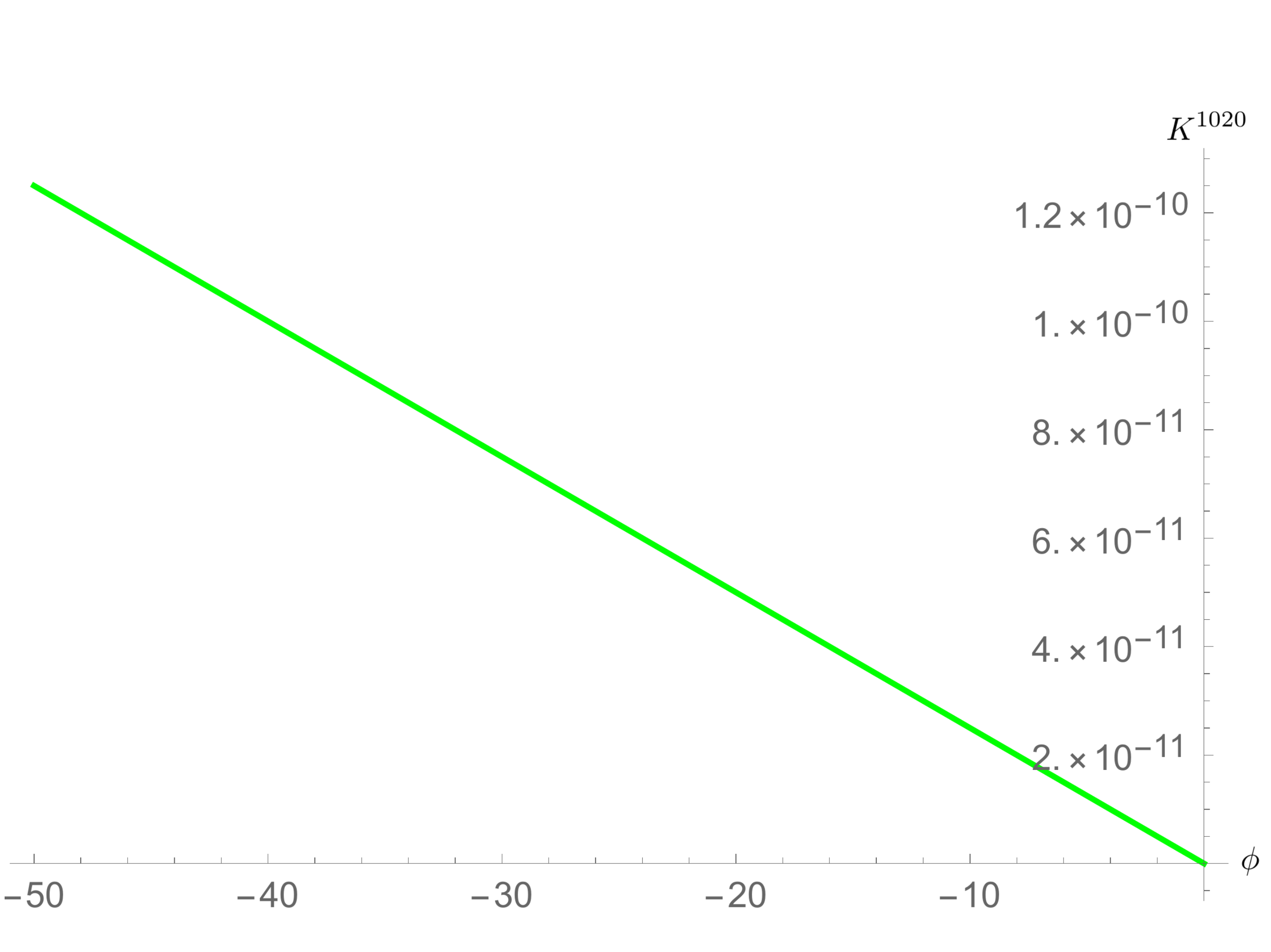}
		\caption{The evolution of the moments $K^{1012}$ and $K^{1020}$ for $\gamma_0=10^8$ and $p_\beta=10^{-1}$.
As opposed to the case shown in the previous figure, in this case, the isotropic counterpart $G^{10}$
exponentially approaches zero towards
the singularity. Therefore, as can be seen in the plots, the anisotropy decelerates
this approach to zero. The only difference between both moments, $G^{1012}$ and $G^{1020}$,
is that the former will approach zero from negative values, whereas the latter will tend
to zero from positive values.
		}
		\label{fig3}
	\end{center}
\end{figure}

For general moments $G^{ijkl}$ mix both sectors, the evolution slightly
differs from the approximate one. In general the behavior of a specific moment $G^{ijkl}$ is qualitatively the same as its isotropic counterpart, 
either increasing or decreasing depending on the sign of the difference
$(i-j)$, but some of them are slightly accelerated or decelerated. The
different behavior does not seem to follow any specific rule based on the
values of the indices. Some examples of such corrections are shown in the
plots depicted in Figs.~\ref{fig1}--\ref{fig3}, where the evolution of some
relative moments $K^{ijkl}$ are shown. The study clearly shows how the
presence of anisotropy affects the evolution of the moments. In all the
cases we observe that, instead of a polynomial of order $k$, the moments
follow a linear dependence in time, that is, $K^{ijkl}=c^{ijkl}\phi$ with
constants $c^{ijkl}$.

There might be several reasons for such a disagreement. On the one hand, the
approximate analytical solution could be invalid because one or several
anharmonicity parameters might not be negligible. On the other hand, it might
well happen that closer to the singularity, where this formalism ceases to be
valid, one recovers the commented polynomial behavior.  Finally, the choice of
peaked states could in principle play a relevant role in the behavior of the
system, but we have tested that this is not the case. If one allows for a
squeezed state, either by increasing or decreasing the value of the Gaussian
widths, $\sigma_v$ and $\sigma_\beta$, moments depart a little bit more from
their corresponding isotropic behavior. But in all cases, deviations from
their isotropic counterparts stay small during the whole evolution.

In summary, for uncorrelated Gaussian initial states we have found that pure
fluctuations of $\beta$ ($K^{00n0}$) and $p_\beta$ ($K^{00n0}$) are conserved
quantities, while the remaining moments $K^{ijkl}$ are not stabilized to any
specific constant values as they approach the singularity. In fact they
diverge linearly in internal time $\phi$, either to plus or minus infinity.

\subsection{General anisotropy}

We now present our extension of the numerical study to the case of a general anisotropy. We
have systematically studied different ranges of values for all the parameters
involved in the evolution in order to understand the global behavior. In
particular, we have found a qualitative change in the evolution of the
moments, and in the approach of the system to the singularity, depending on the value of
$\gamma_0$. Apart from this, another important variable is $p_{\beta}$, which measures the
departure from the harmonic behavior.  As always, for the Hamiltonian
\eqref{classicalH} to be real, the relation $p_{\beta}< \gamma_0$ must hold,
indicating a relationship between these two scales. The relation between
$p_{\beta}$ and $p_{v_0}$ or $v_0$, by contrast, appears to be irrelevant
for the qualitative physical behavior of the system.

Accordingly, this subsection is divided into three parts. In the first two we
analyze the behavior of the moments for different ranges of values of
$\gamma_0$, whereas in the last one, the effects of the quantum back-reaction
on classical trajectories are studied.

\begin{figure}[t]
	\begin{center}
		\includegraphics[scale=0.26]{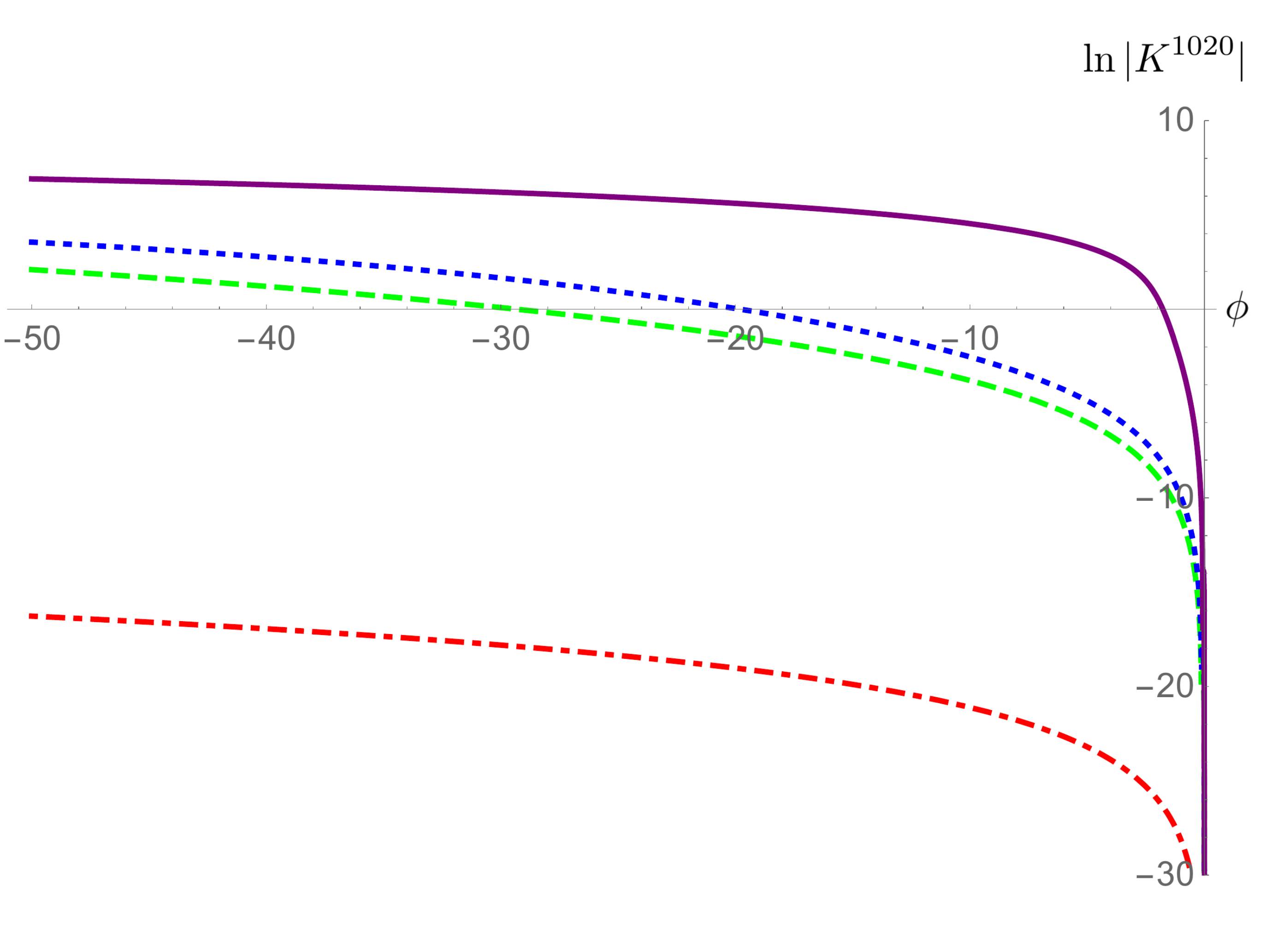}
		\includegraphics[scale=0.26]{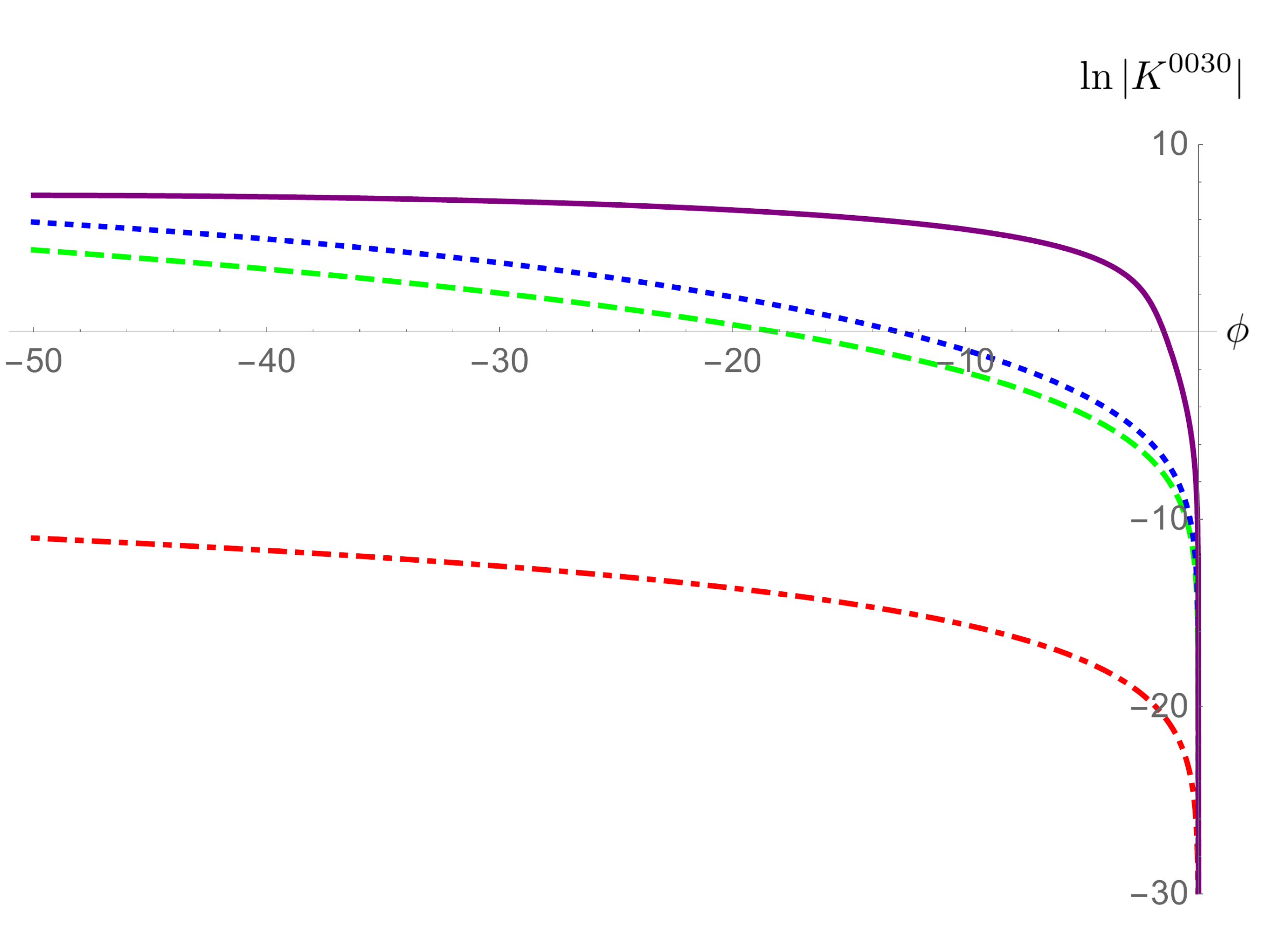}
		\includegraphics[scale=0.257]{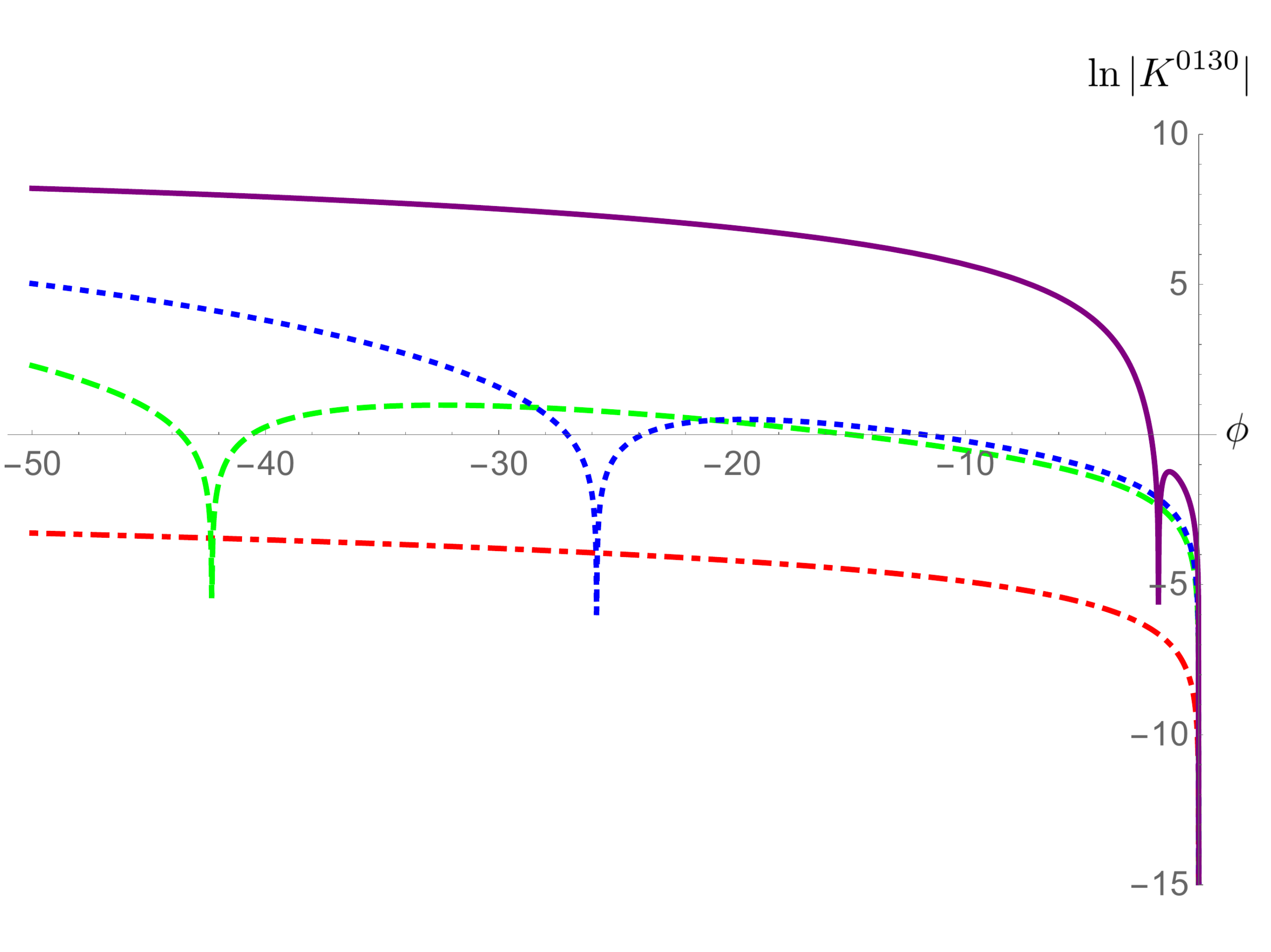}
		\caption{Examples of evolution of dominant moments of the form $K^{00m0}$, $K^{n0m0}$, and
		$K^{0lm0}$ for a relatively small value of $\gamma_0\sim 10^{20}$ in logarithmic plots.
		Different colors and dashing correspond to different values of the ratio $p_\beta/\gamma_0$.
		In particular, purple and continuous $(0.3)$, blue and dotted $(10^{-2})$, green and dashed $(10^{-8})$, red and dot-dashed $(10^{-16})$. Every sharp peak that appears in the last plot, corresponds to a change of sign of the moment. Note that despite those changes the absolute value keeps diverging with the same tendency.
               }
		\label{figcurv}
	\end{center}
\end{figure}

\subsubsection{Evolution of the moments for small values of $\gamma_0$}

We first turn to the behavior of the model for $\gamma_0 \lesssim 10^{20}$.
In this regime, we have analyzed the increase of $p_{\beta}$ approaching its
upper bound given by $\gamma_0$.  As commented above, in order to construct the formalism under
consideration, we have introduced a truncation, assuming that sixth- and higher-order
moments are negligible. It is expected that this approximation holds only as
long as the state is sufficiently peaked around the classical trajectory. As
one departs from the harmonic case the numerical solutions might eventually
break down, signaling the limited validity of the approximation. Due to such
limitations, in this case, it was not possible for us to consider values of $p_\beta$
greater than $0.3\gamma_0$.

While we depart from the limiting case of slowly varying approximation,
more and more moments begin to deviate from their harmonic behavior. None of
the relative moments $K^{ijkl}$ stabilize their behavior; they either increase
or decrease continually towards the singularity. It is interesting to note that
when we depart from the previous regime, we see a certain dilation in the
evolution of moments for a given amount of scalar-field time.  That is, the
evolution of a given moment for a large value of $p_\beta$ during a short
period of time corresponds exactly to the whole evolution of the same moment
for a small value of $p_\beta$ but a longer period of time. This result
indicates some scaling in time, parametrized by $p_\beta$.  Therefore, this
variable drives the velocity of evolution of the moments, in much the same way
that it controls the velocity of the anisotropy, even though it is not
canonically conjugate to the moments.  This effect can be seen especially in the
last plot depicted in Fig.~\ref{figcurv}, as the value of the ratio $p_\beta/\gamma$ increases, the change of sign occurs at earlier times.

Concerning the moments, there is a subset that dominate the dynamics
in the sense that they evolve faster than the other ones, diverging
exponentially towards the singularity, and get a larger absolute value than
the rest of the moments.  In particular the most dominant moments are the pure
fluctuations of $\beta$ ($K^{00m0}$). The other relevant moments are the
correlations between $\beta$ and $v$ ($K^{n0m0}$) and between $\beta$ and
$p_v$ ($K^{0lm0}$). All the mentioned moments (with the particular exception of $K^{0130}$
for certain values of $p_\beta$)
are increasing for even values
of the index $m$, corresponding to $\beta$, and decreasing for odd
values. But this rule does not apply to other generic moments $K^{ijkl}$. In fact, some
of them, depending on the value of $p_\beta$, diverge to minus or plus
infinity, as the commented $K^{0130}$. Nonetheless, in the following section (when the value of $\gamma_0$ is larger than
the one considered here) we will see that almost all moments will follow this rule.

On the contrary, pure fluctuations of $p_v$ ($K^{0n00}$) evolve very slowly,
and they keep a small value along the whole evolution. Therefore, these are
the least affected moments by the presence of the anisotropy.

\subsubsection{Evolution of the moments for large values of $\gamma_0$}

For large values of $\gamma_0$, approximately in the range
$10^{21}$--$10^{25}$, the isotropic dynamics completely dominates the behavior
of different moments. Even for large values of $p_\beta$, up to $0.4\gamma_0$,
the evolution of the moments follow exactly the isotropic one and all
$K^{ijkl}$ are constant. This is also the case if one squeezes the state by
modifying the relation $\sigma_v=\sqrt{\hbar}$ by several orders of
magnitude. The regime of large $\gamma_0$ might also be interpreted as the
classical limit of the model since it implies a very large value of the
classical part of the Hamiltonian, which then dominates over all moment terms.

Contrary to the previous case where the maximum value we could consider for
$p_\beta$ was found to be around $0.3\gamma_0$, here we can choose values as large
as $0.7\gamma_0$. This result is consistent with the smallness of moment terms
relative to the classical contribution to the Hamiltonian, such that
truncation effects should be negligible.  Only for values of $p_{\beta}$
larger than $0.4\gamma_0$ do the moments depart from their corresponding
slowly varying anisotropic behavior. Classical trajectories are then modified
by quantum back-reaction, as will be shown in the next subsection.

For this case, we observe that more moments depart from their harmonic behavior than in
the previous (small $\gamma_0$) case. For such moments, 
we have been able to find a general rule that characterizes their divergence
when approaching the singularity. The key parameter is the moment
index that refers to the shape-parameter $\beta$.  More precisely, as we
approach the singularity, moments $K^{nm(2l)m}$ with an even index in $\beta$
are increasing functions, diverging to positive infinity, whereas moments
$K^{nm(2l+1)m}$ with an odd index in $\beta$ are decreasing functions.  This
rule agrees with what we found in the previous section for a certain subset of
moments, but here it is obeyed by almost all activated moments, with a few exceptions.

\begin{figure}[]
	\begin{center}
		\includegraphics[scale=0.26]{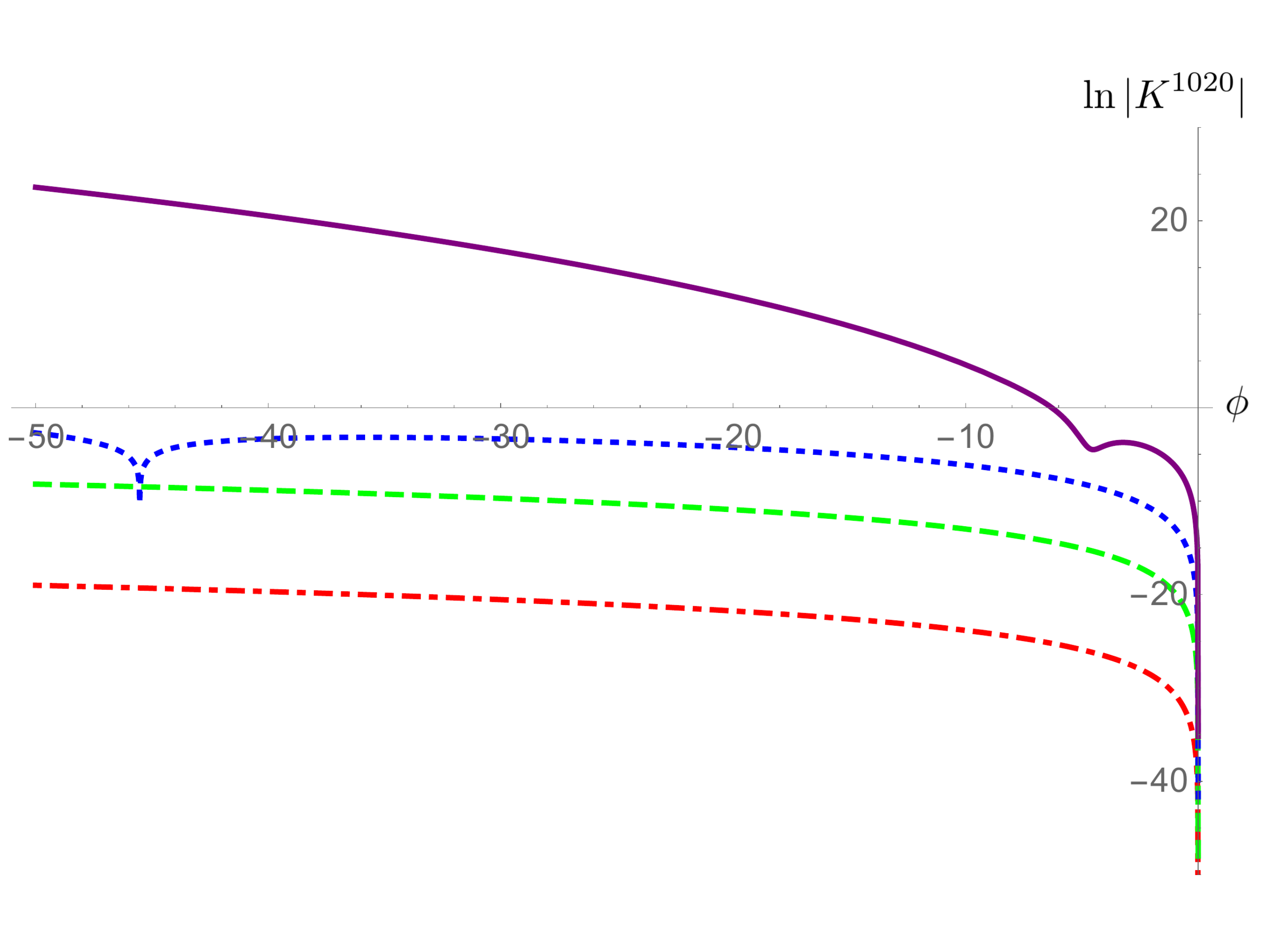}
		\includegraphics[scale=0.26]{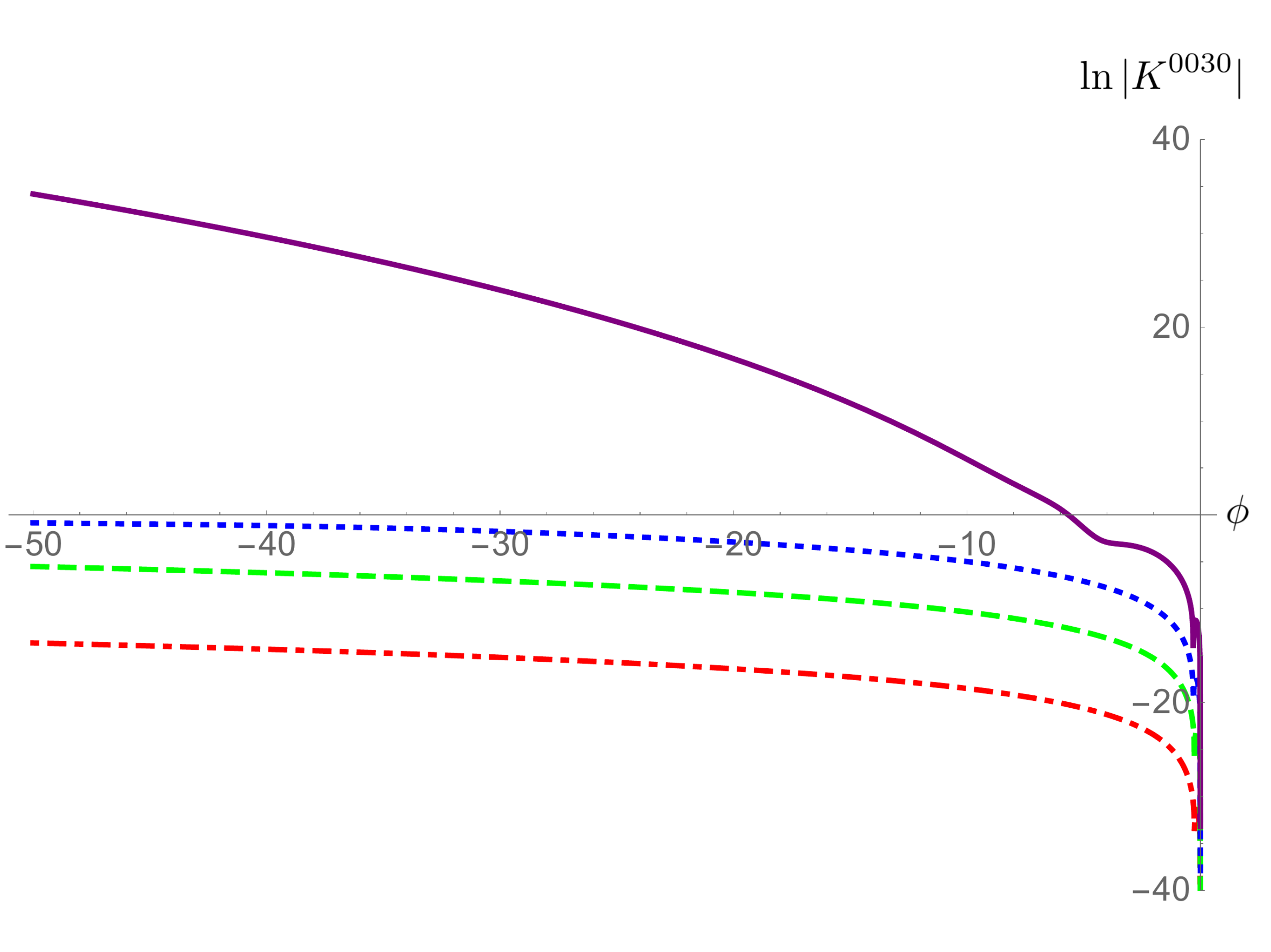}
		\includegraphics[scale=0.26]{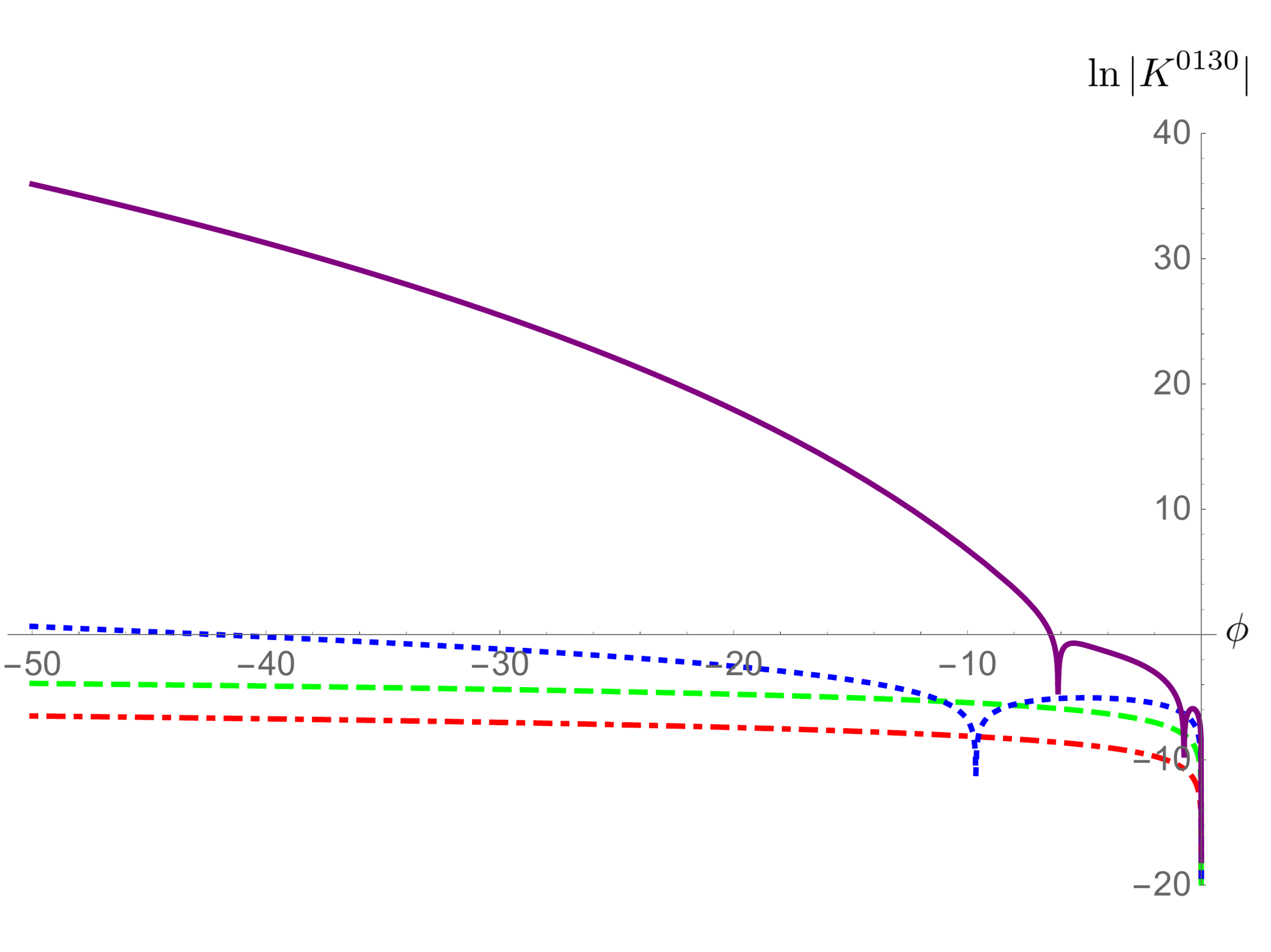}
		\caption{Evolution of the same moments as in
                  Fig.~(\ref{figcurv}) for a larger value of $\gamma_0 \sim
                  10^{21}$. In this case we have chosen the following $p_\beta/\gamma_0$:
                  $0.7$ (purple and continuous line), $0.3$ (blue and dotted line), $10^{-2}$ (green and dashed line) and $10^{-8}$ (red and dot-dashed line).}
		\label{mom-alpha}
	\end{center}
\end{figure}

The absolute value the moments reach at the end of their evolution is much
greater in this case than in the previous one. Furthermore, moments,
which in the previous case were approximately constant, are now evolving.
This outcome
is not related to the fact that here we have been able to get closer to the
limit of $p_\beta \approx \gamma_0$: Even here, the moments are completely
constant if we use the maximum value of the ratio $p_\beta/\gamma_0$ considered before
(around $0.3$), while for larger ratios the moments start increasing much faster than
in the previous case. In order to compare them, we show the evolution of the same moments as in the previous section in Fig.~\ref{mom-alpha}.

Finally, for extremely large values of $\gamma_0\gtrsim 10^{25}$, the
previously mentioned effects are absent and all the $K^{ijkl}$ follow an
almost constant evolution even when we reach the limit $p_\beta\approx \gamma_0$.
This limiting case is relevant because
it shows that the dilation effects in evolution cannot be explained simply by
a $p_{\phi}$ enlarged by moment terms, which would rescale any
$\phi$-derivative in the equations of motion. If this were the reason for dilation
effects, it should occur even in the case of very large $\gamma_0$, in
particular for $p_\beta\approx \gamma_0$ which implies that $p_{\phi}$ is
tiny, leading to large rescaling factors of $1/p_{\phi}$.

\subsubsection{Quantum modifications of the classical trajectories}

The evolution of the expectation values $\beta$ and $v$ is generically
modified by quantum back-reaction effects. Nonetheless, in the approximation
of slowly varying anisotropy we have analytically found, and numerically
checked, that these parameters follow their classical trajectories up to a high
degree of precision.  But in the case of rapidly varying anisotropy, $p_\beta
\approx\gamma_0$, we do observe a departure from the classical behavior.

\begin{figure}[]
	\begin{center}
		\includegraphics[scale=0.3]{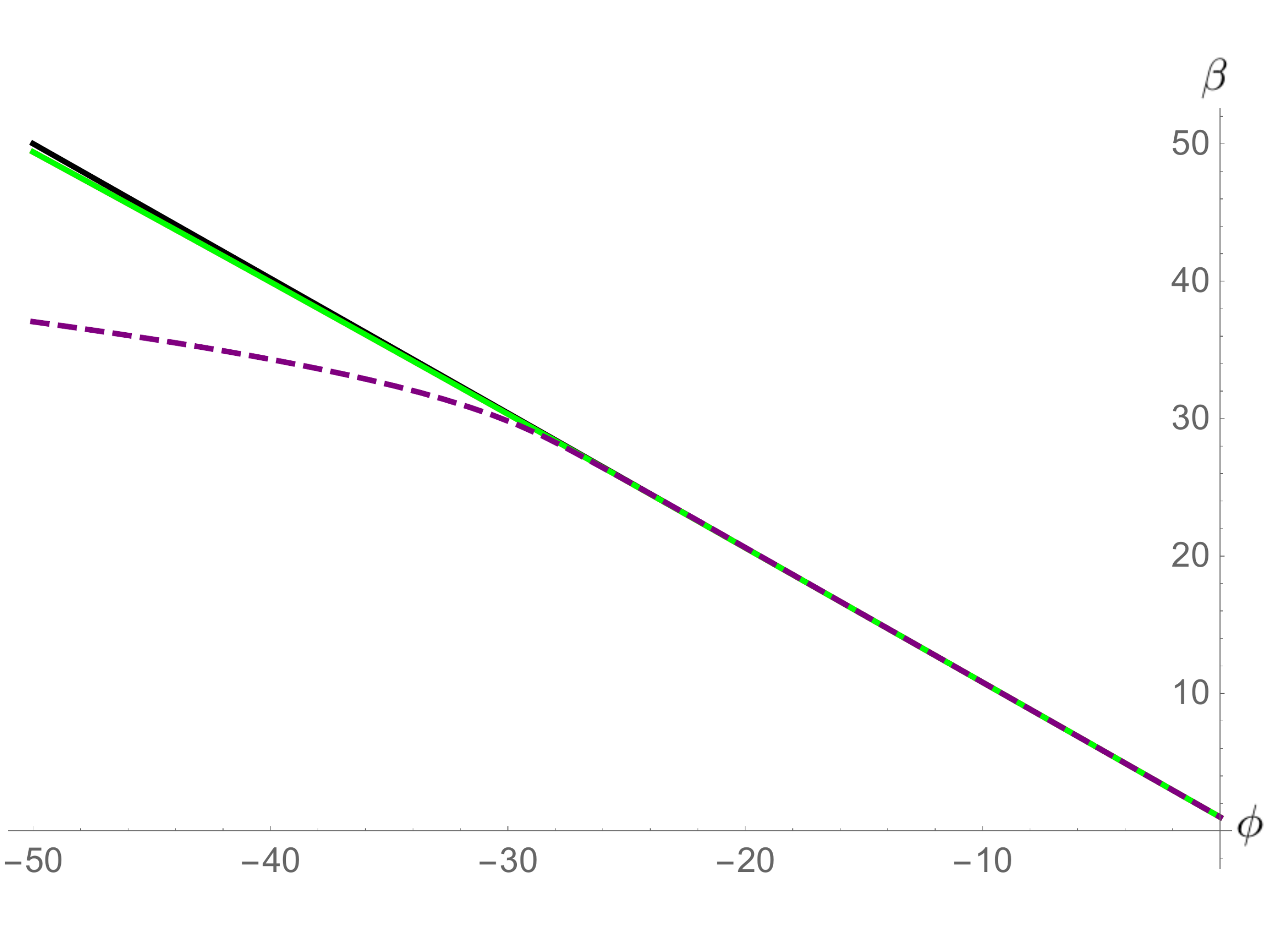}
		\caption{In this plot one can see how the
                  quantum corrections produce a decrease in the growth of the
                  anisotropy close to the singularity for $10^{21}\lesssim\gamma_0\lesssim 10^{25}$.
		Different colors correspond to different truncations in moments.
		 On the one hand, black (classical solution), red (second-order) and green (third-order) are almost overlapping continuous lines.
		On the other hand, the fourth-order (purple and dashed)  line shows the commented slow-down of the anisotropization.
		}
		\label{beta}
	\end{center}
\end{figure}

In the classical setting, the evolution of the shape-parameter $\beta$ is a
linearly increasing function of time, viewed towards the singularity.  However,
as seen in Fig.~\ref{beta}, for large values of $\gamma_0$, quantum effects
give rise to a decrease of the anisotropization toward the
singularity. That is, quantum modifications decelerate the divergent behavior
of the shape-parameter towards the singularity.

\begin{figure}[]
	\begin{center}
		\includegraphics[scale=0.3]{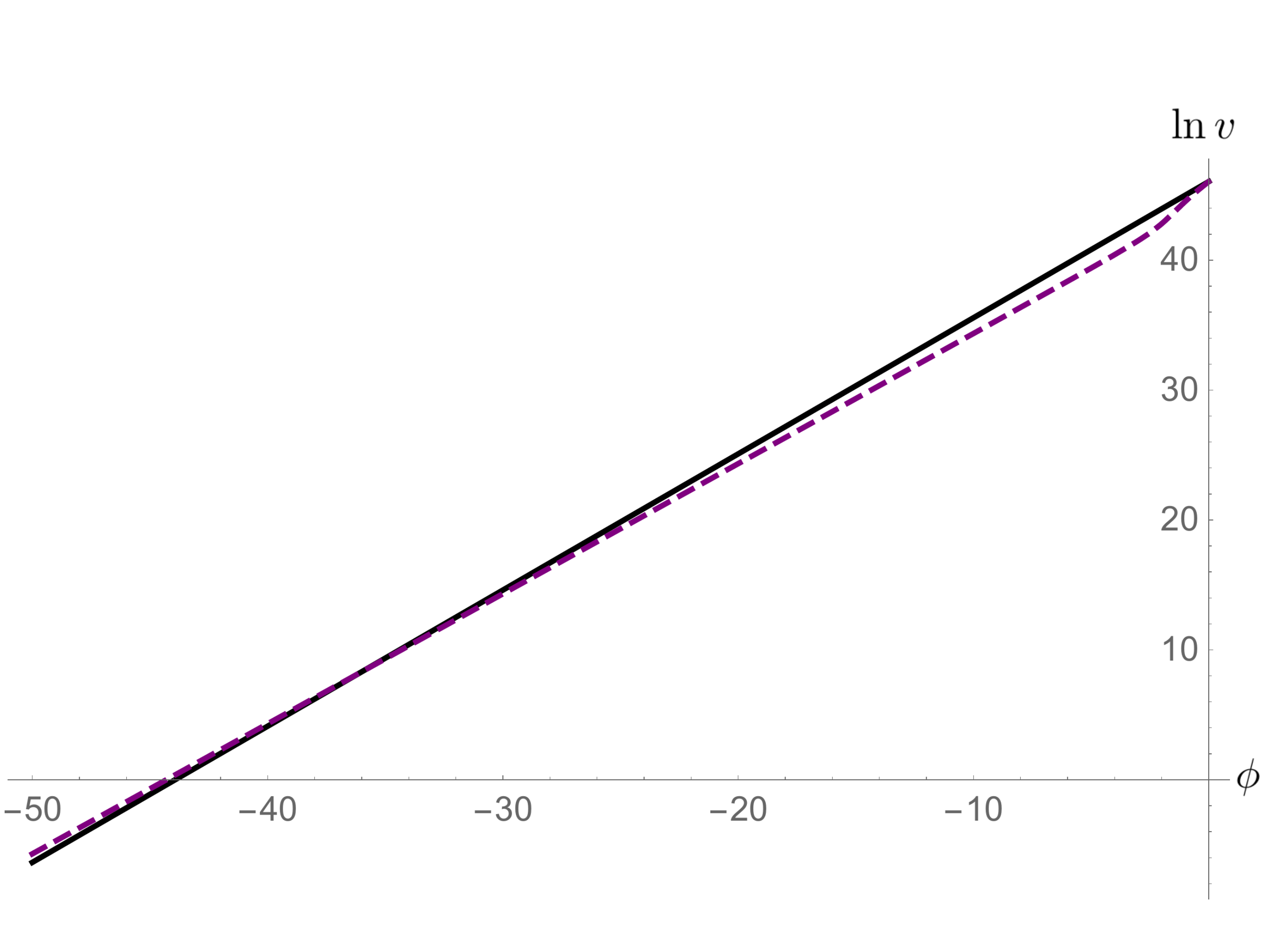}
		\includegraphics[scale=0.3]{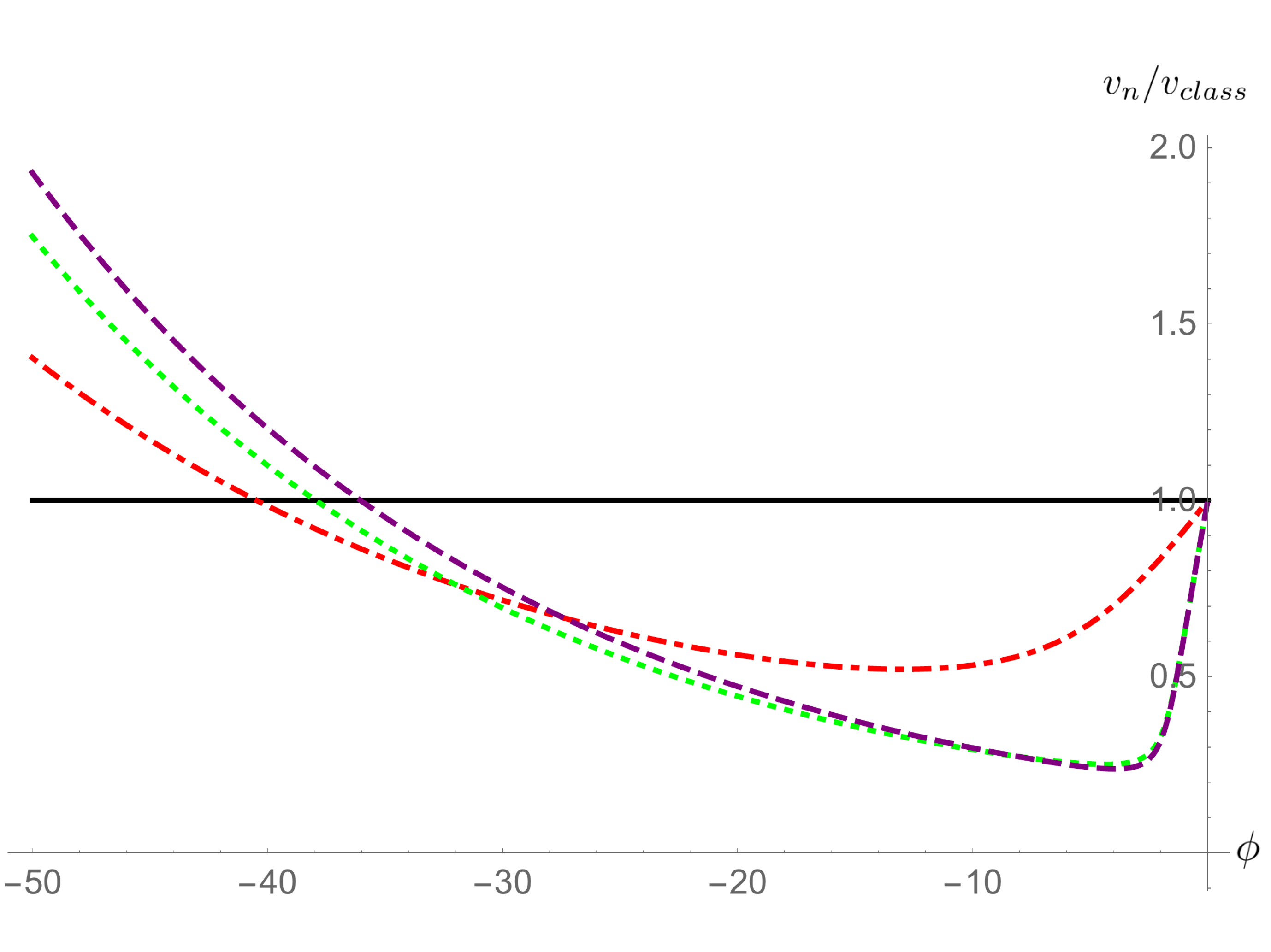}
		\caption{In the first plot we represent the approach of the volume to the
                  singularity in a logarithmic scale for $\gamma_0\sim 10^{20}$. For the sake of clarity, we only represent the classical evolution of the volume (black and continuous line) and the solution truncated at fifth-order (purple and dashed line). In the second plot, in order to show the differences among distinct truncations of the quantum solution, we show the ratio of the volume truncated at $n$-th order ($v_n$) with respect to the classical volume ($v_{class}$). The reference line is represented by the continuous black line, whereas the second and third-order truncations exactly overlap (represented by the red and dot-dashed line). The more relevant effect appears at the fourth-order (given by the green and dotted line) and fifth-order (represented by the purple and dashed line), showing clearly that, even if initially the collapse is faster than in the classical case, it is slower at the final stages of the evolution.}
		\label{fig-vol-fast}
	\end{center}
\end{figure}

In the evolution of the volume $v$, which classically follows an exponentially
decreasing behavior towards the singularity, we observe three qualitatively
different behaviors in the presence of quantum effects: For relatively small values of
$\gamma_0$, $\gamma_0\lesssim 10^{20}$, the volume collapses faster at the
beginning of the evolution, and then follows the standard exponential
behavior but with a slightly lower slope. Thus, it approaches the
singularity slower in the presence of quantum corrections, as can be seen in Fig.~\ref{fig-vol-fast}. The early quantum
modification, appearing as a ``jump'' in the volume, is due to the fact that
the chosen initial state is not a coherent state of the model. Therefore,
initially vanishing moments are turned on, reaching their ``natural'' value
adapted to this particular dynamics. This transition, which has also been seen
at late times in other models \cite{HigherMoments}, produces the fast but
short initial collapse. Once the moments settle down to a more coherent
behavior, we observe the usual exponential collapse of the volume. In this
case, the small value of $\gamma_0$ is not strong enough to hide these quantum
back-reaction effects at an early epoch of the evolution.

\begin{figure}[]
	\begin{center}
		\includegraphics[scale=0.3]{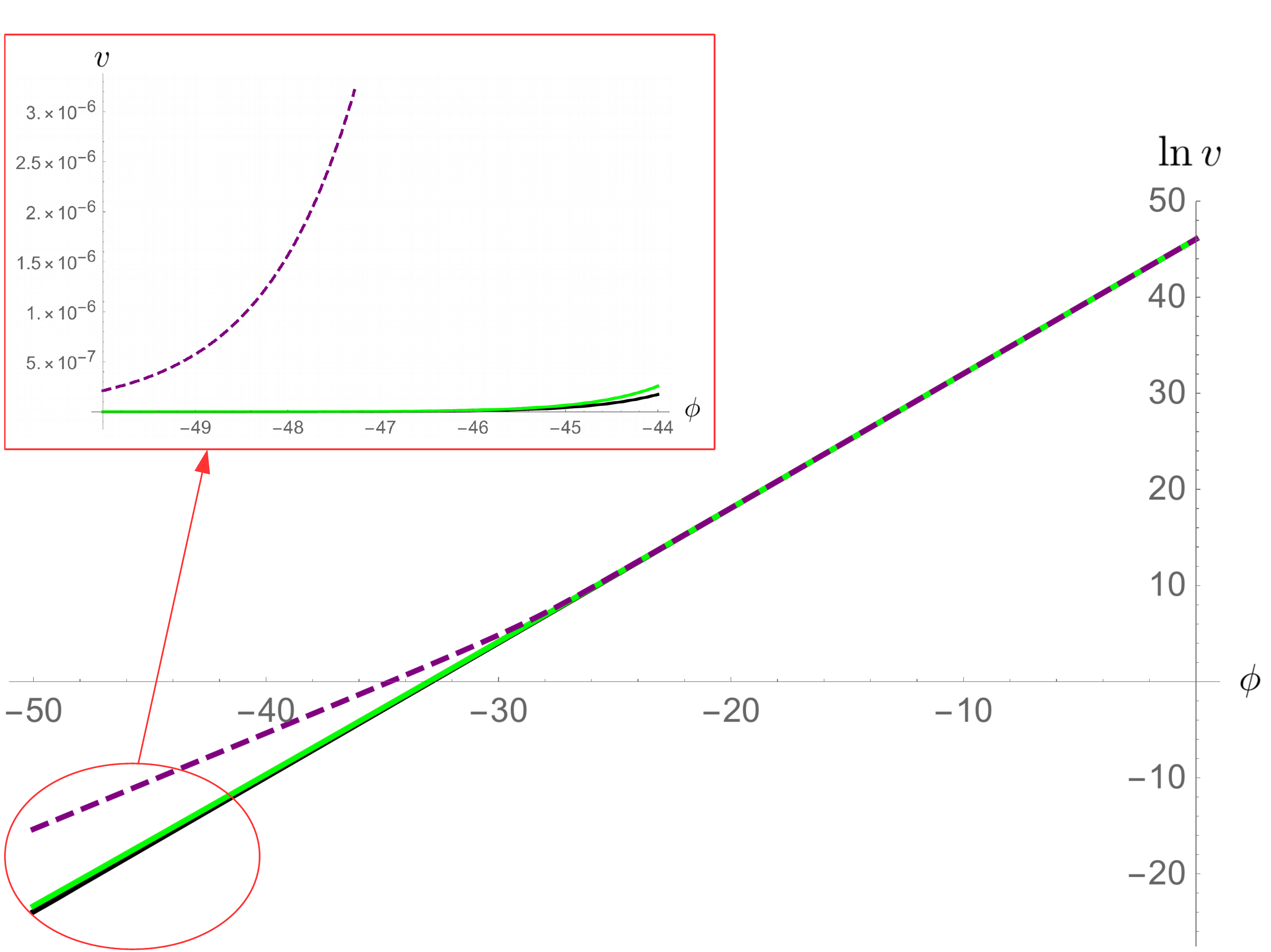}
		\caption{The logarithm of the volume approaching the singularity for $10^{21}\lesssim\gamma_0\lesssim 10^{25}$.
		 The zoom-in shows (in a linear plot) the last
                  stages of the evolution of the volume towards the singularity.
                  Different colors and dashing correspond to different truncations in moments. On the one hand, black (classical solution), red (second-order) and green (third-order) are almost overlapping continuous lines.
		On the other hand, at fourth-order (purple and dashed line), the collapse of the volume is clearly
                  slowed down by quantum effects.}
		\label{vol}
	\end{center}
\end{figure}

Correspondingly, for greater values of $\gamma_0$,
$10^{21}\lesssim\gamma_0\lesssim 10^{25}$, we do not observe the initial
``jump'' in the volume. In that case, quantum corrections lead to a significant slow-down
of the rate of collapse of the volume towards the singularity, as seen in
Fig.~\ref{vol}. This result is consistent with the softening in the increase
of shape-parameter towards the singularity, and points to a smoothening of
the singularity by quantum effects.
\footnote{We would like to point out that a previous study in the context of quantum geometrodynamics also suggests
an avoidance of the classical singularity \cite{Kie19}.}
(We would like to note that for this range of $\gamma_0$ the fifth-order
truncation of the system has shown strong numerical instabilities and, therefore,
the commented result has been derived from the system truncated up to fourth order.)

Finally, as explained previously, very large values of $\gamma_0$, $\gamma_0\gtrsim
10^{25}$, correspond to the classical limit of the model. For this case,
quantum effects are not visible in our numerical implementation, and
expectation values follow exactly their classical trajectories.  In fact, due
to the large value of the classical Hamiltonian, one would expect that quantum
modifications of these trajectories will appear only much closer to the
singularity.

\section{Conclusions}

We have considered a Bianchi I model written in terms of Misner-type
variables. For the sake of simplicity, but without loss of generality, we have
taken only one anisotropic direction to perform our analysis. The classical
model contains a singularity located at a vanishing value of the volume $v=0$
. In this paper we have studied the quantum evolution of the system when
approaching this singularity. With additional approximations the model is
simple enough to allow analytical statements about generic moments, yet
complex enough to show non-trivial behavior.

For such a purpose, we have made use of a formalism based on a decomposition
of a quantum state into its infinite set of moments.  Furthermore, we have
considered a set of variables that allow us to decouple the quantum evolution
for some of the relevant variables, simplifying both the analysis and
interpretation. Because of this choice, we have been able to find analytical
solutions for all moments when the anisotropy evolves slowly (that is, when
the conjugate momentum of the shape-parameter $p_\beta$ is much smaller than
the Hamiltonian $\gamma$ of the reference isotropic model) by performing an
expansion in $p_{\beta}/\gamma$. In this limiting case we can compare with the
isotropic solution ($p_\beta=0$), showing that the evolution of the volume is
not affected. But those moments that in the isotropic case were constant now feel the
anisotropy and evolve, having a polynomial dependence on time. We remark that the
moments with vanishing index on $\beta$ play a special role in the evolution,
being constants of motion.

Going beyond the previous approximation by including terms of arbitrary powers
in $\gamma^{-1}$, we have analyzed the particular scenario when, at an initial
time, the shape-parameter $\beta$ and its momentum $p_{\beta}$ are zero. In
this case the universe is initially isotropic, but one would expect the
generation of an anisotropy during quantum evolution, owing to the performed
classical reduction of symmetry that is not consistent in the quantum
realm. (Note that, based on the uncertainty principle, not all moments can
vanish.) Nonetheless, we have found an analytic solution and shown that
anisotropy is not necessarily generated during quantum evolution. This
interesting new result could be a consequence of the completely decoupled
nature of degrees of freedom in the classical Hamiltonian, which may allow a
classical symmetry to be unaltered by quantization. Nevertheless, it suggests
an interesting middle ground between having no symmetry-preserving solutions
at all (a common criticism of minisuperspace quantization) and not generating
any non-symmetric degrees of freedom (a concern sometimes voiced about
cosmological structure formation).

In order to go beyond the limiting case, we have performed a numerical
simulation of the complete system up to fifth order in moments. We have used
the limiting case to check our method and study the evolution of the moments
towards the singularity. After that, we have considered a general case and
analyzed the evolution for the different ranges of $p_{\beta}$ and the initial
value of $\gamma$. We have shown how some moments are activated during
evolution, departing from their isotropic behavior and diverging as they
approach the singularity. The greater the value of $p_{\beta}$ with respect to
$\gamma_0$ the more dominant the moments, mainly those that represent pure
fluctuations in $\beta$. In addition, they are not only more dominant for that
range, but also allow us to see a longer stretch of their evolution.

Concerning the expectation values, we have analyzed the quantum evolution of
the volume $v$ and the shape-parameter $\beta$, which are determined by
previous variables and moments. We have shown that quantum back-reaction has a
more relevant effect when the initial value of $\gamma$ is bigger. It then
appreciably acts only in a region near the singularity, where we expect
quantum effects even if the initial state is classical. In that case, the
shape-parameter, which classically increases towards the singularity,
decelerates this divergent behavior, softening the anisotropization close to
the singularity. In a similar sense, the exponential collapse of volume
towards the singularity has been smoothed by quantum back-reaction effects,
decreasing the rate of collapse.

Finally, we would like to note that
several statements about the generic moment orders relied on the
$\beta$-independence of the classical Hamiltonian, which would no longer be
the case in models with an anisotropy potential. In such models, it would be
more difficult to obtain corresponding statements, even if a similar behavior
might still be realized. One would then have to rely on numerical
investigations based on the methods developed here. Such results, however,
would always be state dependent and require a careful analysis of implications
of how one chooses an initial state.


\section*{Acknowledgments}
DB gratefully acknowledges hospitality from the
Max-Planck-Institut f\"ur Gravitationsphysik (Albert-Einstein-Institut),
where part of this work has been performed,
and especially the members of the Theoretical Cosmology group. 
AA-S is funded by the Alexander von Humboldt Foundation. Her  work is also
partially supported by the Project. No. MINECO FIS2017-86497-C2-2-P from
Spain. MB was supported in part by NSF grant PHY-1912168.
DB acknowledges funding from the Spanish Ministry of Science for
a stay in foreign research centers,
Project FIS2017-85076-P (MINECO/AEI/FEDER, UE),
and Basque Government Grant No.~IT956-16.


\onecolumngrid

\appendix


\section{Second-order equations of motion for absolute moments}\label{app_eq2G}

In order to give an idea about the system of equations, we present the
second-order truncation of equations of motion, used up to fifth order in the
numerics. 
The equations of motion for the variables
$\{v,p_v,\beta, p_\beta\}$ are
\begin{eqnarray*}
	\dot{v}&=&\frac{1}{H^5}\bigg[ H^4 v^2 p_v+\frac{3}{2}
	v^4 p_v p_\beta^2 G^{0 2 0 0}-v^2 p_\beta \left(p_\beta^2+2 p_v^2 v^2\right) G^{0 1 0 1}
	- vp_v p_\beta \left(2 p_\beta^2+p_v^2 v^2\right) G^{1 0 0 1},
	\\
	&+& v^2p_v\left(\frac{p_v^2 v^2}{2}+ p_\beta^2\right)
	G^{0 0 0 2}+v  p_\beta^2 \left(2 p_\beta^2+p_v^2 v^2\right)
	G^{1 1 0 0}+p_v p_\beta^2\left(p_\beta^2+\frac{1}{2} p_v^2 v^2 \right) G^{2 0 0 0}\bigg] ,
	\\
	\dot p_v &=&\frac{1}{H^5}\bigg[- H^4 v p_v^2-\frac{3}{2} v p_v^4 p_\beta^2 G^{2 0 0 0}
	-\frac{1}{2} v  p_v^2 \left(2 p_\beta^2+p_v^2 v^2\right) G^{0 0 0 2}+p_\beta  p_v^2\left(p_\beta^2+2
	p_v^2 v^2\right) G^{1 0 0 1},
	\\
	&-&p_\beta^2  p_v \left(2 p_\beta^2+p_v^2 v^2\right) G^{1 1 0 0}
	+v p_v p_\beta\left(2 p_\beta^2+ v^2 p_v^2 \right) G^{0 1 0 1}
	-\frac{1}{2} v p_\beta^2 \left(2 p_\beta^2+ v^2p_v^2\right) G^{0 2 0 0}\bigg] ,
	\\
	\dot\beta &=&\frac{1}{H^5}\bigg[-H^4 p_\beta-\frac{3}{2} v^2 p_v^2 p_\beta G^{0 0 0 2}-\frac{1}{2}
	p_v^2p_\beta \left(p_\beta^2+2  v^2 p_v^2\right) G^{2 0 0 0}
	-v  p_v p_\beta \left(2 p_\beta^2+ v^2 p_v^2\right) G^{1 1 0 0},
	\\
	&+&v^2 p_v\left( v^2p_v^2+2 p_\beta^2\right)
	G^{0 1 0 1}-\frac{1}{2}  v^2 p_\beta \left(p_\beta^2+2 v^2 p_v^2 \right) G^{0 2 0 0}+
	v p_v^2\left(v^2 p_v^2+2 p_\beta^2 \right) G^{1 0 0 1}\bigg] ,
	\\
	\dot p_\beta&=&0 .
\end{eqnarray*}
The equations for second-order moments are 
\begin{eqnarray*}
	\dot{G}^{0 0 0 2}&=&0 ,
	\\
	\dot{G}^{0 0 1 1}&=&\frac{v p_v}{H^3}\bigg[ p_v p_\beta G^{1 0 0 1}+ v p_\beta G^{0 1 0 1}-vp_v G^{0 0 0 2}\bigg] ,
	\\
	\dot{G}^{0 0 2 0}&=&\frac{v p_v}{H^3}\bigg[2 p_v p_\beta G^{1 0 1 0}+2  v p_\beta G^{0 1 1 0}-2 v p_v G^{0 0 1 1}\bigg] ,
	\\
	\dot{G}^{0 1 0 1}&=&\frac{p_v}{H^3}\bigg[ - v p_vp_\beta G^{0 0 0 2}+ p_vp_\beta^2 G^{1 0 0 1}+v\left(2 p_\beta^2- v^2p_v^2\right) G^{0 1 0 1}\bigg] ,
	\\
	\dot{G}^{0 1 1 0}&=&\frac{p_v}{H^3}\bigg[- v p_v p_\beta G^{0 0 1 1}
	-v^2 p_v G^{0 1 0 1}+ p_v p_\beta^2 G^{1 0 1 0}+ v p_v p_\beta G^{1 1 0 0}+ v^2 p_\beta G^{0 2 0 0}+ v\left(2 p_\beta^2- v^2 p_v^2\right) G^{0 1 1 0}\bigg] ,
	\\
	\dot{G}^{0 2 0 0}&=&\frac{p_v}{H^3}\bigg[-2 v p_v p_\beta G^{0 1 0 1}
	+2  p_v p_\beta^2 G^{1 1 0 0}+2 v\left(2 p_\beta^2- v^2p_v^2 \right)G^{0 2 0 0}\bigg] ,
	\\
	\dot{G}^{1 0 0 1}&=&\frac{v}{H^3}\bigg[v p_v p_\beta G^{0 0 0 2}
	-v p_\beta^2 G^{0 1 0 1} + p_v \left( v^2 p_v^2-2 p_\beta^2 \right) G^{1 0 0 1}\bigg] ,
	\\
	\dot{G}^{1 0 1 0}&=&\frac{v}{H^3}\bigg[p_v^2 p_\beta G^{2 0 0 0} -v  p_v^2 G^{1 0 0 1}
	+v p_v p_\beta  G^{0 0 1 1}+v p_v p_\beta  G^{1 1 0 0}-v  p_\beta^2 G^{0 1 1 0}
	+ p_v \left( v^2 p_v^2-2 p_\beta^2\right) G^{1 0 1 0}\bigg] ,\\
	\dot{G}^{1 1 0 0}&=&\frac{1}{H^3}\bigg[p_v^2 p_\beta^2 G^{2 0 0 0}- v p_v^2 p_\beta G^{1 0 0 1}
	+ v^2 p_v p_\beta G^{0 1 0 1}- v^2 p_\beta^2 G^{0 2 0 0}\bigg] ,
	\\
	\dot{G}^{2 0 0 0}&=&\frac{2 v}{H^3}\bigg[ v p_v p_\beta G^{1 0 0 1}- v p_\beta^2 G^{1 1 0 0}
	+p_v \left( v^2 p_v^2-2  p_\beta^2\right) G^{2 0 0 0}\bigg] .
\end{eqnarray*}

\section{The equation for the volume at fifth order in moments}\label{app_eq5}
{\small
\begin{eqnarray*}
	\dot v&=&-\frac{3 p_\beta^2 \left(p_\beta^4+12 \gamma ^2 p_\beta^2+8 \gamma
		^4\right) G^{0 5 0 0} v^6}{8 H^{11}}+\frac{5 \left(3 \gamma  p_\beta^4+4 \gamma ^3 p_\beta^2\right)
		G^{0 4 0 0} v^5}{8 H^9}+\frac{5 p_\beta \gamma  \left(15 p_\beta^4+40 \gamma ^2 p_\beta^2+8
		\gamma ^4\right) G^{0 4 0 1} v^5}{8 H^{11}}\\
	&-&\frac{\left(p_\beta^4+4 \gamma ^2 p_\beta^2\right)
		G^{0 3 0 0} v^4}{2 H^7}-\frac{p_\beta \left(3 p_\beta^4+24 \gamma ^2 p_\beta^2+8 \gamma
		^4\right) G^{0 3 0 1} v^4}{2 H^9}-\frac{\left(12 p_\beta^6+159 \gamma ^2 p_\beta^4+136 \gamma ^4
		p_\beta^2+8 \gamma ^6\right) G^{0 3 0 2} v^4}{4 H^{11}}\\&-&\frac{5 p_\beta^2 \gamma  \left(18
		p_\beta^4+41 \gamma ^2 p_\beta^2+4 \gamma ^4\right) G^{1 4 0 0} v^4}{8 H^{11}}+\frac{3 p_\beta^2 \gamma  G^{0 2 0 0} v^3}{2 H^5}+\frac{3 \left(3 \gamma  p_\beta^3+2 \gamma ^3 p_\beta\right)
		G^{0 2 0 1} v^3}{2 H^7}\\&+&\frac{3 \left(2 \gamma ^5+21 p_\beta^2 \gamma ^3+12 p_\beta^4 \gamma \right)
		G^{0 2 0 2} v^3}{4 H^9}+\frac{15 p_\beta \gamma  \left(4 p_\beta^4+13 \gamma ^2 p_\beta^2+4
		\gamma ^4\right) G^{0 2 0 3} v^3}{4 H^{11}}\\&+&\frac{p_\beta^2 \left(4 p_\beta^4+27 \gamma ^2
		p_\beta^2+4 \gamma ^4\right) G^{1 3 0 0} v^3}{2 H^9}+\frac{p_\beta \left(12 p_\beta^6+159
		\gamma ^2 p_\beta^4+136 \gamma ^4 p_\beta^2+8 \gamma ^6\right) G^{1 3 0 1} v^3}{2
		H^{11}}\\&-&\frac{p_\beta \left(p_\beta^2+2 \gamma ^2\right) G^{0 1 0 1} v^2}{H^5}-\frac{\left(2
		p_\beta^4+11 \gamma ^2 p_\beta^2+2 \gamma ^4\right) G^{0 1 0 2} v^2}{2 H^7}-\frac{p_\beta
		\left(2 p_\beta^4+21 \gamma ^2 p_\beta^2+12 \gamma ^4\right) G^{0 1 0 3} v^2}{2 H^9}\\&-&\frac{\left(8
		p_\beta^6+136 \gamma ^2 p_\beta^4+159 \gamma ^4 p_\beta^2+12 \gamma ^6\right) G^{0 1 0 4}
		v^2}{8 H^{11}}-\frac{3 p_\beta^2 \gamma  \left(4 p_\beta^2+\gamma ^2\right) G^{1 2 0 0} v^2}{2
		H^7}\\&-&\frac{3 p_\beta \gamma  \left(12 p_\beta^4+21 \gamma ^2 p_\beta^2+2 \gamma ^4\right)
		G^{1 2 0 1} v^2}{2 H^9}-\frac{3 \gamma  \left(48 p_\beta^6+186 \gamma ^2 p_\beta^4+79 \gamma ^4
		p_\beta^2+2 \gamma ^6\right) G^{1 2 0 2} v^2}{4 H^{11}}\\&-&\frac{p_\beta^2 \left(12 p_\beta^6+159 \gamma ^2 p_\beta^4+136 \gamma ^4 p_\beta^2+8 \gamma ^6\right) G^{2 3 0 0} v^2}{4
		H^{11}}+\frac{\gamma  v}{H}+\frac{\left(\gamma ^3+2 p_\beta^2 \gamma \right) G^{0 0 0 2} v}{2
		H^5}+\frac{\left(2 \gamma  p_\beta^3+3 \gamma ^3 p_\beta\right) G^{0 0 0 3} v}{2 H^7}\\&+&\frac{\gamma 
		\left(8 p_\beta^4+24 \gamma ^2 p_\beta^2+3 \gamma ^4\right) G^{0 0 0 4} v}{8 H^9}+\frac{
		p_\beta \gamma  \left(8 p_\beta^4+40 \gamma ^2 p_\beta^2+15 \gamma ^4\right) G^{0 0 0 5} v}{8
		H^{11}}+\frac{p_\beta^2 \left(2 p_\beta^2+\gamma ^2\right) G^{1 1 0 0} v}{H^5}\\&+&\frac{p_\beta
		\left(2 p_\beta^4+11 \gamma ^2 p_\beta^2+2 \gamma ^4\right) G^{1 1 0 1} v}{H^7}+\frac{\left(4
		p_\beta^6+54 \gamma ^2 p_\beta^4+45 \gamma ^4 p_\beta^2+2 \gamma ^6\right) G^{1 1 0 2} v}{2
		H^9}\\&+&\frac{p_\beta \left(4 p_\beta^6+98 \gamma ^2 p_\beta^4+177 \gamma ^4 p_\beta^2+36 \gamma ^6\right) G^{1 1 0 3} v}{2 H^{11}}+\frac{3 \left(12 \gamma  p_\beta^6+21 \gamma ^3
		p_\beta^4+2 \gamma ^5 p_\beta^2\right) G^{2 2 0 0} v}{4 H^9}\\&+&\frac{3 p_\beta \gamma 
		\left(36 p_\beta^6+177 \gamma ^2 p_\beta^4+98 \gamma ^4 p_\beta^2+4 \gamma ^6\right)
		G^{2 2 0 1} v}{4 H^{11}}-\frac{p_\beta \gamma  \left(2 p_\beta^2+\gamma ^2\right)
		G^{1 0 0 1}}{H^5}-\frac{\left(\gamma ^5+10 p_\beta^2 \gamma ^3+4 p_\beta^4 \gamma \right)
		G^{1 0 0 2}}{2 H^7}\\&-&\frac{\left(4 \gamma  p_\beta^5+22 \gamma ^3 p_\beta^3+9 \gamma ^5
		p_\beta\right) G^{1 0 0 3}}{2 H^9}-\frac{\left(9 \gamma ^7+138 p_\beta^2 \gamma ^5+152 p_\beta^4 \gamma
		^3+16 p_\beta^6 \gamma \right) G^{1 0 0 4}}{8 H^{11}}\\&-&\frac{\left(2 p_\beta^6+11 \gamma ^2
		p_\beta^4+2 \gamma ^4 p_\beta^2\right) G^{2 1 0 0}}{2 H^7}-\frac{\left(2 p_\beta^7+45 \gamma
		^2 p_\beta^5+54 \gamma ^4 p_\beta^3+4 \gamma ^6 p_\beta\right) G^{2 1 0 1}}{2
		H^9}\\&-&\frac{\left(4 p_\beta^8+194 \gamma ^2 p_\beta^6+549 \gamma ^4 p_\beta^4+194 \gamma ^6
		p_\beta^2+4 \gamma ^8\right) G^{2 1 0 2}}{4 H^{11}}-\frac{3 \left(8 \gamma  p_\beta^8+56 \gamma ^3
		p_\beta^6+39 \gamma ^5 p_\beta^4+2 \gamma ^7 p_\beta^2\right) G^{3 2 0 0}}{4
		H^{11}}\\&+&\frac{p_\beta^2 \gamma  \left(2 p_\beta^2+\gamma ^2\right) G^{2 0 0 0}}{2 H^5
		v}+\frac{\left(2 \gamma  p_\beta^5+11 \gamma ^3 p_\beta^3+2 \gamma ^5 p_\beta\right)
		G^{2 0 0 1}}{2 H^7 v}+\frac{\left(2 \gamma ^7+45 p_\beta^2 \gamma ^5+54 p_\beta^4 \gamma ^3+4
		p_\beta^6 \gamma \right) G^{2 0 0 2}}{4 H^9 v}\\&+&\frac{\left(4 \gamma  p_\beta^7+98 \gamma ^3
		p_\beta^5+177 \gamma ^5 p_\beta^3+36 \gamma ^7 p_\beta\right) G^{2 0 0 3}}{4 H^{11}
		v}+\frac{\left(12 \gamma ^2 p_\beta^6+21 \gamma ^4 p_\beta^4+2 \gamma ^6 p_\beta^2\right)
		G^{3 1 0 0}}{2 H^9 v}\\&+&\frac{\left(4 p_\beta \gamma ^8+98 p_\beta^3 \gamma ^6+177 p_\beta^5
		\gamma ^4+36 p_\beta^7 \gamma ^2\right) G^{3 1 0 1}}{2 H^{11} v}-\frac{p_\beta^2 \gamma ^3 \left(4
		p_\beta^2+\gamma ^2\right) G^{3 0 0 0}}{2 H^7 v^2}\\&-&\frac{\left(2 p_\beta \gamma ^7+21
		p_\beta^3 \gamma ^5+12 p_\beta^5 \gamma ^3\right) G^{3 0 0 1}}{2 H^9 v^2}-\frac{\left(2 \gamma ^9+79 p_\beta^2 \gamma ^7+186 p_\beta^4 \gamma ^5+48 p_\beta^6 \gamma ^3\right) G^{3 0 0 2}}{4 H^{11}
		v^2}\\&-&\frac{\left(12 \gamma ^2 p_\beta^8+159 \gamma ^4 p_\beta^6+136 \gamma ^6 p_\beta^4+8
		\gamma ^8 p_\beta^2\right) G^{4 1 0 0}}{8 H^{11} v^2}+\frac{\left(4 p_\beta^2 \gamma ^7+27
		p_\beta^4 \gamma ^5+4 p_\beta^6 \gamma ^3\right) G^{4 0 0 0}}{8 H^9 v^3}\\&+&\frac{\left(8 p_\beta \gamma ^9+136 p_\beta^3 \gamma ^7+159 p_\beta^5 \gamma ^5+12 p_\beta^7 \gamma ^3\right)
		G^{4 0 0 1}}{8 H^{11} v^3}-\frac{\left(4 p_\beta^2 \gamma ^9+41 p_\beta^4 \gamma ^7+18 p_\beta^6 \gamma ^5\right) G^{5 0 0 0}}{8 H^{11} v^4}.
\end{eqnarray*}}
\newpage
\section{Second-order equations of motion for relative moments}\label{app_eq2K}

Performing the change of variables $\{v,p_v,\beta,p_\beta,G^{ijkl}\}$
to $\{v,\gamma,\beta,p_\beta,K^{ijkl}\}$,
the equations of motion for $\{\gamma, p_\beta,K^{ij0l}\}$
decouple from the rest, $p_\beta$ and all its fluctuations $K^{000n}$ are
constants of motion. 

\begin{eqnarray*}
	\dot{\gamma} &=&\frac{\gamma ^2 p_\beta}{H^3} \left(p_\beta
	(K^{0 2 0 0}-K^{2 0 0 0})-K^{0 1 0 1}+K^{1 0 0 1}\right), \\
	\dot{p_\beta}&=&0, \\
	\dot{K}^{0 0 0 2}&=&0, \\
	\dot{K}^{0 1 0 1}&=&\frac{\gamma  p_\beta}{H^3} \left(p_\beta (K^{0 1 0 1}+K^{1 0 0 1})-K^{0 0 0 2}
	\right), \\ \dot{K}^{0 2 0 0}&=&\frac{2 \gamma  p_\beta}{H^3}
	\left(p_\beta (K^{0 2 0 0}+K^{1 1 0 0})-K^{0 1 0 1}
	\right), \\ \dot{K}^{1 0 0 1}&=&-\frac{\gamma  p_\beta}{H^3} \left(p_\beta (K^{0 1 0 1}+K^{1 0 0 1})-K^{0 0 0 2}
	\right), \\  \dot{K}^{1 1 0 0}&=&-\frac{\gamma  p_\beta}{H^3}
	\left(p_\beta (K^{0 2 0 0}-K^{2 0 0 0})-K^{0 1 0 1}
	+K^{1 0 0 1}\right), \\ \dot{K}^{2 0 0 0}&=&-\frac{2 \gamma  p_\beta}{H^3} \left(p_\beta (K^{1 1 0 0}+K^{2 0 0 0})-K^{1 0 0 1}
	\right).
\end{eqnarray*}

The moments $K^{ijkl}$ related to $\beta$ are given by

\begin{eqnarray*}
	\dot{K}^{0 0 1 1}&=&\frac{\gamma ^2}{H^3} \left(p_\beta (K^{0 1 0 1}+K^{1 0 0 1}
	)-K^{0 0 0 2}\right), \\
	\dot{K}^{0 0 2 0}&=&\frac{2 \gamma ^2}{H^3} \left(p_\beta (K^{0 1 1 0}+K^{1 0 1 0})-K^{0 0 1 1}
	\right), \\ \dot{K}^{0 1 1 0}&=&\frac{\gamma}{H^3}  \left(-\gamma  K^{0 1 0 1}
	+p_\beta (\gamma  (K^{0 2 0 0}+K^{1 1 0 0}
	)-K^{0 0 1 1})+p_\beta^2 (K^{0 1 1 0}+K^{1 0 1 0}
	)\right), \\ \dot{K}^{1 0 1 0}&=&-\frac{\gamma}{H^3}  \left(\gamma  K^{1 0 0 1}
	-p_\beta (\gamma  (K^{1 1 0 0}+K^{2 0 0 0}
	)+K^{0 0 1 1})+p_\beta^2 (K^{0 1 1 0}+K^{1 0 1 0}
	)\right). \\ 
\end{eqnarray*}

Just as these equations, the equation for $\beta$ is also independent of the
volume:
\begin{eqnarray*}
	\dot{\beta}&=&-\frac{1}{2 H^5}\bigg(-2 \gamma ^4 (K^{0 1 0 1}+K^{1 0 0 1})+\gamma^2 p_\beta^3 (K^{0 2 0 0}+4 K^{1 1 0 0}+K^{2 0 0 0}
	-4)
	) \\
	&&  -4 \gamma ^2 p_\beta^2 (K^{0 1 0 1}+K^{1 0 0 1}+ p_\beta \left(2 \gamma ^4 (K^{0 2 0 0}+K^{1 1 0 0}
	+K^{2 0 0 0}+1)+3 \gamma ^2 K^{0 0 0 2}\right)+2 p_\beta^5\bigg).
\end{eqnarray*}

Finally, the equation of motion for the volume is given by a logarithmic
derivative,
\begin{eqnarray*}
	\frac{\dot{v}}{v}&=&\frac{\gamma}{2 H^5} \bigg(2 \gamma^4
	-4 \gamma^2 p_\beta^2+2p_\beta^4
	+\gamma ^2 K^{0 0 0 2}+p_\beta^2 \left(\gamma ^2 (3
	K^{0 2 0 0}+2 K^{1 1 0 0}+K^{2 0 0 0})+2 K^{0 0 0 2}\right)\nonumber\\
	&-&2
	\gamma ^2 p_\beta (2 K^{0 1 0 1}+K^{1 0 0 1})+2
	p_\beta^4 (2 K^{1 1 0 0}+K^{2 0 0 0})-2 p_\beta^3 (K^{0 1 0 1}+2 K^{1 0 0 1})\bigg).
\end{eqnarray*}

\newpage
\section{The solution for the quasiharmonic case truncated at third order in moments}\label{app_sol}

In this appendix we present
the solution for the quasiharmonic case up to order ${\cal O}(2)$ and truncated
at third order in moments. For the generic case, with $r\neq 0$,
moments $K^{ijkl}$ go as $\phi^{k/2}$ and, in particular moments of the form
$K^{ij0l}$ are constants of motion. The remaining moments are
\begin{eqnarray*}
	K^{0 0 1 1}&=& c_1+\frac{2 \sqrt{\gamma_0^2+r \phi }}{r}
	(-K^{0 0 0 2}+K^{0 1 0 2}+K^{1 0 0 2}+p_\beta
	(K^{0 1 0 1}-K^{0 2 0 1}+K^{1 0 0 1}-K^{1 1 0 1}-K^{2 0 0 1})), \\ K^{0 0 1 2}&=&
	c_2+\frac{2 \sqrt{\gamma_0^2+r \phi }}{r} (p_\beta
	(K^{0 1 0 2}+K^{1 0 0 2})-K^{0 0 0 3}), \\ K^{0 1 1 0}&=& c_3+\frac{2
		\sqrt{\gamma_0^2+r \phi }}{r}
	(-K^{0 1 0 1}+K^{0 2 0 1}+K^{1 1 0 1}+p_\beta
	(K^{0 2 0 0}-K^{0 3 0 0}+K^{1 1 0 0}-K^{1 2 0 0}-K^{2 1 0 0})), \\ K^{0 1 1 1}&=&
	c_4+\frac{2 \sqrt{\gamma_0^2+r \phi }}{r} (p_\beta
	(K^{0 2 0 1}+K^{1 1 0 1})-K^{0 1 0 2}), \\ K^{0 2 1 0}&=& c_5+\frac{2
		\sqrt{\gamma_0^2+r \phi }}{r} (p_\beta
	(K^{0 3 0 0}+K^{1 2 0 0})-K^{0 2 0 1}), \\ K^{1 0 1 0}&=& c_6+\frac{2
		\sqrt{\gamma_0^2+r \phi }}{r}
	(-K^{1 0 0 1}+K^{1 1 0 1}+K^{2 0 0 1}+p_\beta
	(K^{1 1 0 0}-K^{1 2 0 0}+K^{2 0 0 0}-K^{2 1 0 0}-K^{3 0 0 0})), \\ K^{1 0 1 1}&=&
	c_7+\frac{2 \sqrt{\gamma_0^2+r \phi }}{r} \bigg(p_\beta
	(K^{1 1 0 1}+K^{2 0 0 1})-K^{1 0 0 2}\bigg), \\ K^{0 0 2 1}&=& c_8-\frac{4
		\sqrt{\gamma_0^2+r \phi }}{r} (c_2-p_\beta c_4-p_\beta
	c_7)\\ &&+\frac{4 \phi}{r}  \bigg(K^{0 0 0 3}+p_\beta (-2 K^{0 1 0 2}-2
	K^{1 0 0 2}+p_\beta (K^{0 2 0 1}+2
	K^{1 1 0 1}+K^{2 0 0 1}))\bigg), \\ 
	K^{1 1 1 0}&=& c_9+\frac{2 \sqrt{\gamma_0^2+r \phi }}{r} (p_\beta
	(K^{1 2 0 0}+K^{2 1 0 0})-K^{1 1 0 1}), \\ K^{0 1 2 0}&=& c_{10}-\frac{4
		\sqrt{\gamma_0^2+r \phi }}{r} (c_4-p_\beta c_5-p_\beta
	c_9) \\&&+\frac{4 \phi}{r}  \bigg(K^{0 1 0 2}+p_\beta (-2 K^{0 2 0 1}-2
	K^{1 1 0 1}+p_\beta (K^{0 3 0 0}+2
	K^{1 2 0 0}+K^{2 1 0 0}))\bigg), \\ K^{2 0 1 0}&=& c_{11}+\frac{2 \sqrt{\gamma_0^2+r \phi }}{r} \bigg(p_\beta
	(K^{2 1 0 0}+K^{3 0 0 0})-K^{2 0 0 1}\bigg), \\ K^{0 0 2 0}&=& c_{12}-\frac{4
		\sqrt{\gamma_0^2+r \phi }}{r} (c_1-p_\beta c_3-c_4+p_\beta c_5-p_\beta c_6-c_7+p_\beta c_9+p_\beta
	c_{11}) \\ && +\frac{4 \phi}{r}  (K^{0 2 0 0}-2 K^{0 3 0 0}+2 K^{1 1 0 0}-4
	K^{1 2 0 0}+K^{2 0 0 0}-4 K^{2 1 0 0}-2 K^{3 0 0 0}) p_\beta^2 \\ &&-\frac{8\phi}{r} \bigg( (K^{0 1 0 1}-2
	K^{0 2 0 1}+K^{1 0 0 1}-3 K^{1 1 0 1}-2 K^{2 0 0 1}) p_\beta+K^{0 0 0 2}-2
	(K^{0 1 0 2}+K^{1 0 0 2})\bigg), \\ K^{1 0 2 0}&=& c_{13}-\frac{4 \sqrt{\gamma_0^2+r \phi }}{r} (c_7-p_\beta c_9-p_\beta
	c_{11}) \\ &&+\frac{4 \phi}{r}  \bigg(K^{1 0 0 2}+p_\beta (-2 K^{1 1 0 1}-2
	K^{2 0 0 1}+p_\beta (K^{1 2 0 0}+2
	K^{2 1 0 0}+K^{3 0 0 0}))\bigg),
\end{eqnarray*}
\newpage
and finally
\begin{eqnarray*}
	K^{0 0 3 0}&=& c_{14}+\frac{6}{r} \bigg(12 \phi 
	(c_2+p_\beta (-2 c_4-2 c_7+p_\beta (c_5+2
	c_9+c_{11})))+\sqrt{\gamma_0^2+r \phi } (p_\beta
	(c_{10}+c_{13})-c_8)\bigg) \\ 
	&& +\frac{12 \gamma_0^2}{r^2}
	\bigg(c_2+p_\beta (-2 
	c_4-2 c_7+p_\beta (c_5+2 c_9+c_{11}))\bigg) \\
	&&-\frac{24 \gamma_0^2 \sqrt{\gamma_0^2+r \phi }}{r^2} p_\beta^3 (K^{0 3 0 0}+3
	K^{1 2 0 0}+3 
	K^{2 1 0 0}+K^{3 0 0 0})\\
	&&-\frac{24 \gamma_0^2 \sqrt{\gamma_0^2+r \phi }}{r^2}p_{\beta}^2
	(K^{0 2 0 1}+2 K^{1 1 0 1}+K^{2 0 0 1}) \\
	&&+\frac{24 \gamma_0^2 \sqrt{\gamma_0^2+r \phi }}{r^3}p_{\beta}(
	K^{0 1 0 2}+K^{1 0 0 2})\\
	&&+\frac{24 \gamma_0^2 \sqrt{\gamma_0^2+r \phi }}{r^3}K^{0 0 0 3}\\
	&& +\frac{8 (\gamma_0^2+r \phi )^{3/2} }{r^3}p_\beta^3  (K^{0 3 0 0}+3
	K^{1 2 0 0}+3 
	K^{2 1 0 0}+K^{3 0 0 0}) \\
	&& -\frac{24 (\gamma_0^2+r \phi )^{3/2} }{r^3} p_\beta^2 
	(K^{0 2 0 1}+2 K^{1 1 0 1}+K^{2 0 0 1}) \\
	&& +\frac{24 (\gamma_0^2+r \phi )^{3/2}}{r^2} p_{\beta}(
	K^{0 1 0 2}+ K^{1 0 0 2})\\
	&&-\frac{8 (\gamma_0^2+r \phi )^{3/2}}{r^2}K^{0 0 0 3}\, ,
\end{eqnarray*}
with integration constants $c_i$.

For the particular case with $r=0$, a moment of the form $K^{ijkl}$ is given
by a polynomial of order $k$ in $\phi$:
\begin{eqnarray*}
	K^{0 0 1 1}&=& c_1+\frac{\phi }{\gamma_0}
	\bigg(-K^{0 0 0 2}+K^{0 1 0 2}+K^{1 0 0 2}+p_\beta
	(K^{0 1 0 1}-K^{0 2 0 1}+K^{1 0 0 1}-K^{1 1 0 1}-K^{2 0 0 1})\bigg), \\ K^{0 0 1 2}&=& \frac{1}{\gamma_0}\bigg(\gamma_0 c_2-\phi  K^{0 0 0 3}+p_\beta \phi 
	(K^{0 1 0 2}+K^{1 0 0 2})\bigg), \\ K^{0 0 2 0}&=& c_3+\frac{ p_\beta^2\phi ^2}{\gamma_0^2}\bigg(K^{0 2 0 0}-2
	K^{0 3 0 0}+2 K^{1 1 0 0}-4 K^{1 2 0 0}+K^{2 0 0 0}-4 K^{2 1 0 0}-2 K^{3 0 0 0}\bigg) \\ &&-\frac{\phi ^2}{\gamma_0^2}\bigg(2 (K^{0 1 0 1}-2 K^{0 2 0 1}+K^{1 0 0 1}-3 K^{1 1 0 1}-2 K^{2 0 0 1}) p_\beta+K^{0 0 0 2}-2 (K^{0 1 0 2}+K^{1 0 0 2})\bigg) \\ &&-\frac{2\phi }{\gamma_0}
	\bigg(c_1-c_7-c_{11}+p_\beta (-c_6+c_9-c_{10}+c_{13}+c_{14})\bigg),  \\ K^{0 0 2 1}&=&c_4+ \frac{\phi
		^2}{\gamma_0^2}\bigg(K^{0 0 0 3}+p_\beta (-2 K^{0 1 0 2}-2
	K^{1 0 0 2}+p_\beta (K^{0 2 0 1}+2 K^{1 1 0 1}+K^{2 0 0 1}))\bigg) \\&& +\frac{2\phi
	}{\gamma_0} \bigg(p_\beta (c_7+c_{11})-c_2\bigg), \\ K^{0 0 3 0}&=& c_5+\frac{p_\beta \phi ^3}{\gamma_0^3}\bigg((K^{0 3 0 0}+3
	K^{1 2 0 0}+3 K^{2 1 0 0}+K^{3 0 0 0}) p_\beta^2-3 (K^{0 2 0 1}+2
	K^{1 1 0 1}+K^{2 0 0 1}) p_\beta\bigg)\\&& +\frac{\phi ^3}{\gamma_0^3}\bigg(p_\beta \left(+3 K^{0 1 0 2}+3
	K^{1 0 0 2}\right)-K^{0 0 0 3}\bigg) +\frac{3\phi
		^2}{\gamma_0^2}
	\bigg(c_2+p_\beta (-2 c_7-2 c_{11}+p_\beta (c_9+2 c_{13}+c_{14}))\bigg) \\&& +\frac{3\phi
	}{\gamma_0} \bigg(p_\beta (c_8+c_{12})-c_4\bigg),
\end{eqnarray*}
\newpage
for pure anisotropy moments, and
\begin{eqnarray*}
 K^{0 1 1 0}&=& c_6+\frac{\phi }{\gamma_0}
	\bigg(-K^{0 1 0 1}+K^{0 2 0 1}+K^{1 1 0 1}+p_\beta
	(K^{0 2 0 0}-K^{0 3 0 0}+K^{1 1 0 0}-K^{1 2 0 0}-K^{2 1 0 0})\bigg), \\ K^{0 1 1 1}&=& \frac{1}{\gamma_0}\bigg(\gamma_0 c_7-\phi  K^{0 1 0 2}+p_\beta \phi 
	(K^{0 2 0 1}+K^{1 1 0 1})\bigg), \\ K^{0 1 2 0}&=&c_8+
	\frac{\phi ^2}{\gamma_0^2}\bigg(K^{0 1 0 2}+p_\beta (-2 K^{0 2 0 1}-2 K^{1 1 0 1}+p_\beta
	(K^{0 3 0 0}+2 K^{1 2 0 0}+K^{2 1 0 0}))\bigg)\\&&+\frac{2\phi }{\gamma_0}
	\bigg(p_\beta (c_9+c_{13})-c_7\bigg), \\ K^{0 2 1 0}&=&
	\frac{1}{\gamma_0}\bigg(\gamma_0 c_9-\phi  K^{0 2 0 1}+p_\beta \phi 
	(K^{0 3 0 0}+K^{1 2 0 0})\bigg), \\ K^{1 0 1 0}&=& c_{10}+\frac{\phi }{\gamma_0}
	\bigg(-K^{1 0 0 1}+K^{1 1 0 1}+K^{2 0 0 1}+p_\beta
	(K^{1 1 0 0}-K^{1 2 0 0}+K^{2 0 0 0}-K^{2 1 0 0}-K^{3 0 0 0})\bigg), \\ K^{1 0 1 1}&=& \frac{1}{\gamma_0}\bigg(\gamma_0 c_{11}-\phi  K^{1 0 0 2}+p_\beta \phi
	(K^{1 1 0 1}+K^{2 0 0 1})\bigg), \\ K^{1 0 2 0}&=&c_{12}+
	\frac{\phi ^2}{\gamma_0^2}\bigg(K^{1 0 0 2}+p_\beta (-2 K^{1 1 0 1}-2 K^{2 0 0 1}+p_\beta
	(K^{1 2 0 0}+2 K^{2 1 0 0}+K^{3 0 0 0}))\bigg) \\&& +\frac{2 \phi }{\gamma_0}
	\bigg(p_\beta (c_{13}+c_{14})-c_{11}\bigg), \\ K^{1 1 1 0}&=&
	\frac{1}{\gamma_0}\bigg(\gamma_0 c_{13}-\phi  K^{1 1 0 1}+p_\beta \phi 
	(K^{1 2 0 0}+K^{2 1 0 0})\bigg), \\ K^{2 0 1 0}&=& \frac{1}{\gamma_0}\bigg(\gamma_0
	c_{14}-\phi  K^{2 0 0 1}+p_\beta \phi  (K^{2 1 0 0}+K^{3 0 0 0})\bigg),
\end{eqnarray*}
for moments with volume-anisotropy correlations.

\section{Third-order solution for the $p_\beta=0$ case}\label{app_solpbeta0}

We present the solution for the system truncated at third order in moments
for the particular case $p_\beta=0$. For the case with $r\neq 0$
the fluctuations of the anisotropic sector are 
\begin{eqnarray*}
	K^{0 0 1 1}&=&c_{1}+ \frac{2(K^{0  0  0  2}-(K^{0  1  0  2}+K^{1  0  0  2}))}{K^{0  1  0  2}-K^{1  0  0  2}}\gamma ,  \qquad
	K^{0 0 1 2}=c_{2}+ \frac{2 K^{0  0  0  3}}{K^{0  1  0  2}-K^{1  0  0  2}}\gamma  , \\
	K^{0 0 2 0}&=&c_{3}+ \frac{4 \gamma}{(K^{0  1  0  2}-K^{1  0  0  2})^2} \bigg[ (K^{0  0  0  2}-2 (K^{0  1  0  2}+K^{1  0  0  2}))\gamma +(K^{0  1  0  2}-K^{1  0  0  2}) (c_{1}+c_{4}+c_{5})\bigg], \\
	K^{0 0 2 1}&=& \frac{1}{(K^{0  1  0  2}-K^{1  0  0  2})^2}\bigg[-8 \gamma_0^2 K^{0  0  0  3}-(K^{0  1  0  2}-K^{1  0  0  2})^2
	c_{6}+ \nonumber \\
	&& 4 (K^{0  1  0  2}-K^{1  0  0  2}) c_{2} \gamma+ 4 K^{0  0  0  3} \gamma ^2\bigg] ,\\
	K^{0 0 3 0}&=& \frac{1}{(K^{0  1  0  2}-K^{1  0  0  2})^3} \bigg[
	(K^{0  1  0  2}-K^{1  0  0  2})^3 c_{7}-24 (K^{0  1  0  2}-K^{1  0  0  2}) \gamma_0^2 c_{2} \nonumber \\
	&& -6 \left(8 \gamma_0^2 K^{0  0  0  3}+ (K^{0  1  0  2}-K^{1  0  0  2})^2 c_{6}\right)\gamma 
	+12(K^{0  1  0  2}-K^{1  0  0  2}) c_{2} \gamma^2 
	+ 8 K^{0  0  0  3}\gamma ^3
	\bigg] ,
\end{eqnarray*}
\newpage
where $c_i$ with $i=1,...,6$ are real constants.
The second-order correlations between the two sectors, with their logarithmic behavior read as:
\begin{eqnarray*}
	K^{0 1 0 1}&=& d_{1}+\frac{K^{0  0  0  3}}{(K^{0  1  0  2}-K^{1  0  0  2})} \ln\gamma, \\
	K^{0 1 1 0}&=&d_{2} -\frac{2}{K^{0  1  0  2}-K^{1  0  0  2}}\bigg(K^{0  2  0  1}-d_{1}+K^{1  1  0  1}\bigg)\gamma 
	+\frac{((K^{0  1  0  2}-K^{1  0  0  2}) c_{2}+2 \gamma  K^{0  0  0  3})}{(K^{0  1  0  2}-K^{1  0  0  2})^2}\ln\gamma ,\\
	K^{1 0 0 1}&=& d_{3}-\frac{K^{0  0  0  3}}{(K^{0  1  0  2}-K^{1  0  0  2})} \ln\gamma ,\\
	K^{1 0 1 0}&=& d_{4}-\frac{2 (-d_{3}+K^{1  1  0  1}+K^{2  0  0  1})}{K^{0  1  0  2}-K^{1  0  0  2}}\gamma
	-\frac{((K^{0  1  0  2}-K^{1  0  0  2}) c_{2}+ 2\gamma  K^{0  0  0  3})}{(K^{0  1  0  2}-K^{1  0  0  2})^2}\ln\gamma ,
\end{eqnarray*}
where $d_i$ with $i=1,...,5$ are real constants.
And, finally, the third-order correlations between the two sectors:
\begin{eqnarray*}
K^{0 1 1 1}&=&-c_{5}+\bigg(1+\frac{K^{0  1  0  2}+K^{1  0  0  2}}{K^{0  1  0  2}-K^{1  0  0  2}}\bigg) \gamma , \\
K^{0 1 2 0}&=&f_{1}+ \frac{4 K^{0  1  0  2}}{(K^{0  1  0  2}-K^{1  0  0  2})^2}\gamma ^2
-\frac{4 c_{5}}{K^{0  1  0  2}-K^{1  0  0  2}} \gamma-\frac{\hbar^2}{2 (K^{0  1  0  2}-K^{1  0  0  2})} \ln\gamma ,\\
K^{0 2 1 0}&=&f_{2}+ \frac{2 K^{0  2  0  1}}{K^{0  1  0  2}-K^{1  0  0  2}}\gamma ,
\\
K^{1 0 1 1}&=&-c_{4}+ \left(\frac{K^{0  1  0  2}+K^{1  0  0  2}}{K^{0  1  0  2}-K^{1  0  0  2}}-1\right)\gamma  ,\\
K^{1 0 2 0}&=&f_{3}+ \frac{4 K^{1  0  0  2}}{(K^{0  1  0  2}-K^{1  0  0  2})^2}\gamma ^2 -\frac{4 
	c_{4}}{K^{0  1  0  2}-K^{1  0  0  2}} \gamma+\frac{\hbar^2}{2 (K^{0  1  0  2}-K^{1  0  0  2})} \ln\gamma ,\\
K^{1 1 1 0}&=& f_{4}+\frac{2 K^{1  1  0  1}}{K^{0  1  0  2}-K^{1  0  0  2}}\gamma  ,\\
K^{2 0 1 0}&=&f_{5}+ \frac{2 K^{2  0  0  1}}{K^{0  1  0  2}-K^{1  0  0  2}}\gamma ,
\end{eqnarray*}
where $f_i$ with $i=1,...,5$ are real constants.

For the case with $r=0$, the solution is slightly different and the fluctuations of the anisotropic sector take the form
\begin{eqnarray*}
	K^{0 0 1 1}&=&\tilde c_{1} -\frac{(K^{0  0  0  2}-K^{0  1  0  2}-K^{1  0  0  2})}{\gamma } \phi , \qquad
	K^{0 0 1 2}=\tilde c_{2}-\frac{K^{0  0  0  3}}{\gamma } \phi , \\
	K^{0 0 2 0}&=& \tilde c_{3}
	-\frac{2 }{\gamma} (\tilde c_{1}+\tilde c_{4}+\tilde c_{5}) \phi
	+\frac{(K^{0  0  0  2}-2 K^{0  1  0  2}-2K^{1  0  0  2}) \phi^2}{\gamma ^2} ,\\
	K^{0 0 2 1}&=& -\tilde c_{6}
	-\frac{2 \tilde c_{2}}{\gamma } \phi
	+\frac{K^{0  0  0  3}}{\gamma ^2} \phi ^2 ,
	\qquad
	K^{0 0 3 0}=\tilde c_{7}
	+\frac{3 \tilde c_{6}}{\gamma} \phi 
	+\frac{3\tilde c_{2}}{\gamma ^2} \phi ^2
	-\frac{K^{0  0  0  3}}{\gamma ^3} \phi ^3.
\end{eqnarray*}
The second-order correlations between the two sectors are given as 
\begin{eqnarray*}
	K^{0 1 0 1}=\tilde d_{1} -\frac{K^{0  0  0  3}}{2 \gamma ^2} \phi , \qquad
	K^{0 1 1 0}=-\tilde d_{2}
	-\frac{  (\tilde c_{2}-2 \gamma (K^{0  2  0  1}+K^{1  1  0  1}-\tilde d_{1}))}{2 \gamma ^2}\phi
	+ \frac{K^{0  0  0  3}}{2 \gamma ^3} \phi ^2 ,
	\\
	K^{1 0 0 1}=\tilde d_{3}+ \frac{K^{0  0  0  3}}{2 \gamma ^2} \phi , \qquad
	K^{1 0 1 0}=\tilde d_{4}
	+\frac{  (2 \gamma (K^{1  1  0  1}+K^{2  0  0  1}-\tilde d_{3})+\tilde c_{2})}{2 \gamma ^2}\phi-\frac{K^{0  0  0  3}}{2 \gamma ^3} \phi ^2 .
\end{eqnarray*}
And finally, the third-order correlations between the two sectors
take the form,
\begin{eqnarray*}
	K^{0 1 1 1}&=&-\tilde c_{5} -\frac{K^{0  1  0  2}+K^{1  0  0  2}}{2 \gamma } \phi , \qquad
	K^{0 1 2 0}= \tilde f_{1}
	+\frac{\left(8 \gamma \tilde c_{5}+\hbar^2\right)}{4 \gamma ^2}\phi  
	+\frac{K^{0  1  0  2}+K^{1  0  0  2}}{2 \gamma ^2} \phi^2 ,
	\\
	K^{0 2 1 0}&=& \tilde f_{2}-\frac{K^{0  2  0  1}}{\gamma } \phi ,\\
	K^{1 0 1 1}&=& -\tilde c_{4} -\frac{K^{0  1  0  2}+K^{1  0  0  2}}{2 \gamma } \phi ,\qquad
	K^{1 0 2 0}= \tilde f_{3}
	-\frac{ \left(\hbar^2-8 \gamma \tilde c_{4}\right)}{4 \gamma ^2}\phi 
	+\frac{K^{0  1  0  2}+K^{1  0  0  2}}{2 \gamma ^2} \phi ^2 ,
	\\
	K^{1 1 1 0}&=& \tilde f_{4}-\frac{K^{1  1  0  1}}{\gamma }\phi  ,\\
	K^{2 0 1 0}&=& \tilde f_{5}-\frac{K^{2  0  0  1}}{\gamma }\phi .
\end{eqnarray*}

\end{document}